\documentclass[%
 reprint,
 superscriptaddress,
 amsmath,amssymb,
 aps,
 prl
]{revtex4-2}

\usepackage{lipsum}  
\usepackage{csquotes} 
\usepackage[caption = false]{subfig} 
\usepackage[breaklinks=true,colorlinks=true,citecolor=blue,urlcolor=blue,linkcolor=blue,pdfencoding=auto, psdextra]{hyperref}

\usepackage{graphicx}
\usepackage{dcolumn}
\usepackage{bm}

\usepackage[normalem]{ulem}
\usepackage{xcolor}

\DeclareMathOperator{\sech}{sech}

\usepackage{bigints} 

\begin{document}

\preprint{APS/123-QED}

\title{Nonthermal Particle Acceleration by Magnetic Pumping in Pulsating Plasmas}

\author{Giuseppe Arrò}
\email{arro@wisc.edu}
\affiliation{Department of Physics, University of Wisconsin-Madison, Madison, Wisconsin 53706, USA}

\author{Vladimir Zhdankin}
\affiliation{Department of Physics, University of Wisconsin-Madison, Madison, Wisconsin 53706, USA}

\author{Fabio Bacchini}
\affiliation{Centre for mathematical Plasma Astrophysics, Department of Mathematics, KU Leuven, Celestijnenlaan 200B, B-3001 Leuven, Belgium}
\affiliation{Royal Belgian Institute for Space Aeronomy, Solar-Terrestrial Centre of Excellence, Ringlaan 3, 1180 Uccle, Belgium}


\begin{abstract}

We present a new ``pulsating box'' setup to investigate particle acceleration in high-$\beta$ plasmas undergoing compression--expansion cycles. Our fully kinetic simulations show that particles are efficiently accelerated by magnetic pumping, producing nonthermal energy distributions with power-law tails. Numerical results are in excellent agreement with a generalized maximum entropy model that we derive, linking the power-law index of distributions to the injected energy. Our results could be relevant for understanding the origin of high-energy particles in certain high-$\beta$ space and astrophysical plasmas.

\end{abstract}


\maketitle

{\it Introduction---}High-energy particles are an important component of many space and astrophysical plasmas, often exhibiting energy distributions with extended power-law tails \citep{mcconnell2002soft,fisk2012particle,oka2018electron,salem2023precision}. The origin of nonthermal particles is debated and several processes have been considered to explain their acceleration. In this context, a key parameter is the ratio of thermal pressure $P$ to magnetic pressure $B^2/(8\pi)$, the plasma beta $\beta\!=\!8\pi P/B^2$. In low-$\beta$ plasmas, such as the solar corona \citep{cranmer2019properties,kasper2021parker} or astrophysical jets of compact objects \citep{janssen2021event,davelaar2023synchrotron}, free energy mainly comes from magnetic fluctuations that can be dissipated through magnetic reconnection \citep{guo2014formation,sironi2014relativistic} and turbulence \citep{zhdankin2017kinetic,comisso2018particle}, accelerating particles. Conversely, magnetic energy is subdominant when $\beta\!\gg\!1$, making reconnection and turbulence inefficient particle accelerators \citep{ball2018electron,comisso2022ion}. However, high-$\beta$ plasmas are often embedded in large-scale subsonic compressible flows (where shocks cannot develop) that could provide energy to accelerate particles. Such systems include planetary magnetosheaths, experiencing compression and expansion under the variable solar wind activity \citep{hellinger2005magnetosheath,travnivcek2007magnetosheath}; accretion disks of compact objects, where gravity, radiation and convection compress and expand the plasma \citep{penna2010simulations,ripperda2020magnetic}; the intracluster medium of galaxy clusters, compressed and expanded by orbital interactions \citep{kunz2011thermally,tran2023electron}.

Compression and expansion cycles increase and decrease the plasma thermal and magnetic energies periodically, producing pressure-anisotropy through conservation of adiabatic invariants. In collisionless plasma, pressure-anisotropy variations are then limited by kinetic instabilities that trap and scatter particles, isotropizing their velocity distribution \citep{gary1991electromagnetic,gary1993theory,kunz2014firehose}. Consequently, part of the energy injected via compression is locked into particle distributions and is not returned during expansion, causing irreversible heating. This mechanism, dubbed {\it magnetic pumping} (MP), has been proposed as a possible source of high-energy particles in space and astrophysical plasmas \citep{berger1958heating,lichko2017magnetic,egedal2021fast,montag2022field,malkov2026magnetic}. Theoretical models show that MP can generate power-law particle distributions consistent with those observed in space and astrophysical plasmas, like the Earth's magnetosheath \citep{lichko2020magnetic}, but supporting numerical evidence is lacking. Furthermore, MP has been studied in plasmas subjected to incompressible shear flows \citep{ley2023heating}, but not in plasmas embedded in compressible flows. Hence, demonstrating that MP produces nonthermal particle distributions in high-$\beta$ plasmas is still an unsolved and largely unexplored problem.

In this Letter, we investigate particle acceleration by MP in high-$\beta$ ``pulsating'' plasmas, i.e.\ plasmas undergoing compression--expansion cycles. Using first-principle simulations implementing a new ``pulsating box'' setup, we provide the first numerical demonstration that MP efficiently produces nonthermal particle distributions with high-energy power-law tails. Our simulations are in excellent agreement with a generalized maximum entropy (GME) model that we derive, describing the universal evolution of the power-law index of particle distributions with the injected energy.

{\it Methods---}We perform particle-in-cell simulations with Zeltron \citep{cerutti2013simulations}, implementing a compressing/expanding box method \citep{sironi2015electron,bacchini2026particle,petersdebonhome2026}. Field and particle equations are solved in a comoving frame where the domain coordinates remain fixed in time. This noninertial change of frame introduces fictitious forces that inject/remove energy from the plasma, accounting for compression/expansion. Lab coordinates $\bm{x}$ and comoving coordinates $\bm{x}^{\prime}$ are related by
\begin{equation}
\bm{x} = \textbf{L}\,\bm{x}^{\prime} =
\begin{pmatrix}
1 & 0 & 0 \\
0 & a & 0 \\
0 & 0 & a
\end{pmatrix}
\bm{x}^{\prime},
\quad\quad a = a_0^{\sin(2\pi\,t/\tau_0)},
\label{matrix}
\end{equation}
describing pulsations with period $\tau_0$ and amplitude $a_0$ in the $yz$-plane, perpendicular to an initially uniform magnetic-field $\bm{B}_0\!=\!(B_0,\,0,\,0)$. We consider a pair plasma, initializing electrons and positrons from uniform Maxwell--J{\"u}ttner distributions \footnote{The Maxwell--J{\"u}ttner distribution is the relativistic generalization of the Maxwell distribution, typically used to initialize particle-in-cell simulations.} \citep{juttner1911maxwellsche} with equal densities $n_0$ and subrelativistic temperatures $\theta_0\!\equiv\!T_0/(m_e c^2)\!=\!0.04$. The initial Alfvén-to-light speed ratio is $v_{A,0}/c\!\equiv\!B_0/\sqrt{8\pi n_0 m_e c^2}\!\simeq\!0.07$, and $\beta_0\!\equiv\!16\pi n_0 T_0/B_0^2\!=\!16$. We employ a 2D periodic domain of initial size $L_x\!=\!L_y\!=\!64\,\rho_{e,0}$ (where $\rho_{e,0}\!\equiv\!\sqrt{\theta_0} m_e c^2/(e B_0)$ is the initial electron gyroradius), sampled by a uniform mesh with $1024^2$ cells, each containing $512$ particles per species. We choose $\tau_0\!=\!400\,\Omega_{e,0}^{-1}$ (where $\Omega_{e,0}\!\equiv\!eB_0/(m_e c)$ is the initial electron gyrofrequency) and run simulations until $t\!=\!10\,\tau_0$ \footnote{We require $\tau_0$ to be larger than time-scales associated with instabilities, comparable to particle gyroperiods.}. Equation~(\ref{matrix}) implies that the box size oscillates between $a_0\,L_y$ and $L_y/a_0$ in the $y$-direction. Pulsations in the $z$-direction are also accounted for by fictitious forces, despite the 2D geometry. Our fiducial run employs $a_0\!=\!0.5$, meaning that the domain in the $y$-direction is compressed to $L_y/2$ during $0\!<\!t/\tau_0\!\leqslant\!0.25$, expands to $2L_y$ during $0.25\!<\!t/\tau_0\!\leqslant\!0.75$, shrinking back to $L_y$ during $0.75\!<\!t/\tau_0\!\leqslant\!1$ (after which the cycle repeats). In real systems, $a_0$ could be estimated by measuring the amplitude of magnetic field and density fluctuations, scaling as $\sim\!1/a^2$ \citep{sironi2015electron}.

\begin{figure}[t]
\centering
\includegraphics[width=0.99\linewidth]{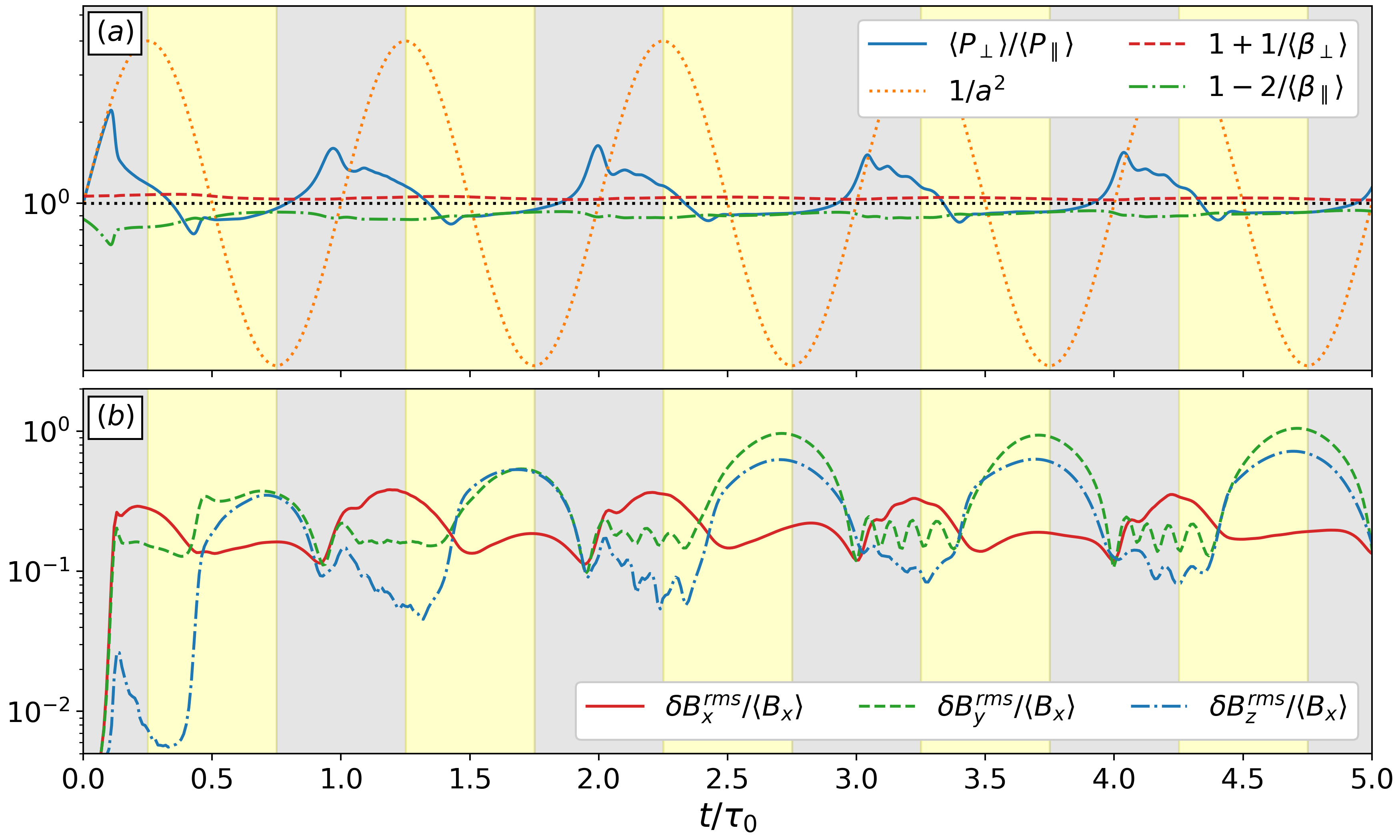}
\caption{Temporal evolution of run $a_0\!=\!0.5$. (a) Pressure-anisotropy (solid-blue), adiabatic scaling (dotted-orange), mirror (dashed-red) and firehose (dash-dotted-green) instability thresholds, $\langle P_\perp \rangle/\langle P_\parallel \rangle \!=\! 1$ (dotted-black). (b) Componentwise root-mean-squared magnetic-field fluctuations. Gray(yellow)-shaded areas highlight compression (expansion) cycles.}
\label{anisotropy}
\end{figure}

{\it Results---}To identify instabilities that mediate MP, we analyze the pressure-anisotropy $\langle P_\perp \rangle/\langle P_\parallel \rangle$, where $P_\parallel\!\equiv\!\textbf{P}\!:\!\bm{B}\bm{B}/B^2$, $P_\perp\!\equiv\!\textbf{P}\!:\!(\textbf{I}-\bm{B}\bm{B}/B^2)/2$, $\textbf{P}$ is the total pressure tensor (including all species, without bulk-flow contributions), $\bm{B}$ is the magnetic-field, and $\langle \cdot \rangle$ denotes volume averaging. Figure~\ref{anisotropy}(a) shows the temporal evolution of $\langle P_\perp \rangle/\langle P_\parallel \rangle$ for run $a_0\!=\!0.5$ (until $t\!=\!5\,\tau_0$), compared to the subrelativistic adiabatic scaling $\langle P_\perp \rangle/\langle P_\parallel \rangle \!=\! 1/a^2$ \citep{sironi2015electron}. After a brief initial adiabatic growth, $\langle P_\perp \rangle/\langle P_\parallel \rangle$ oscillates nonadiabatically within a narrower range than $1/a^2$, limited by the periodic excitation of instabilities. Specifically, $\langle P_\perp \rangle/\langle P_\parallel \rangle$ exceeds the mirror-instability threshold when $\langle P_\perp \rangle/\langle P_\parallel \rangle \!>\! 1+1/\langle \beta_\perp \rangle$ \citep{gary1992mirror,southwood1993mirror}, during compression (gray-shaded areas), and the firehose-instability threshold when $\langle P_\perp \rangle/\langle P_\parallel \rangle \!<\! 1-2/\langle \beta_\parallel \rangle$ \citep{gary1998proton,zhdankin2023synchrotron}, during expansion (yellow-shaded areas), with $\beta_\perp\!=\!8\pi\,P_\perp/B^2$ and $\beta_\parallel\!=\!8\pi\,P_\parallel/B^2$. Correspondingly, magnetic-field fluctuations $\delta \bm{B}\!=\!\bm{B}-\langle\bm{B}\rangle$ consistent with mirror and firehose modes develop when thresholds are crossed. This is illustrated in Fig.~\ref{anisotropy}(b), showing $\delta B_i^{\rm rms}\!=\!\sqrt{\langle \delta B^2_i \rangle}$ (with $i\!=\!x,\,y,\,z$), normalized to $\langle B_x \rangle$. Fluctuations with $|\delta B_x| \!>\! |\delta B_y| \!>\! |\delta B_z|$ develop when the plasma becomes mirror-unstable (in 3D, the mirror-instability would excite modes with spatial variations in $z$ and $|\delta B_x| \!>\! |\delta B_y| \!\sim\! |\delta B_z|$), while fluctuations with $|\delta B_y| \!\sim\! |\delta B_z| \!>\! |\delta B_x|$ emerge when the system becomes firehose-unstable. In addition to mirror modes, $\delta B^{\rm rms}_y$ and $\delta B^{\rm rms}_z$ oscillations develop during compression, corresponding to cyclotron waves \citep{gary1992mirror,lopez2016relativistic}. 

\begin{figure*}[t]
\centering
\includegraphics[width=0.99\linewidth]{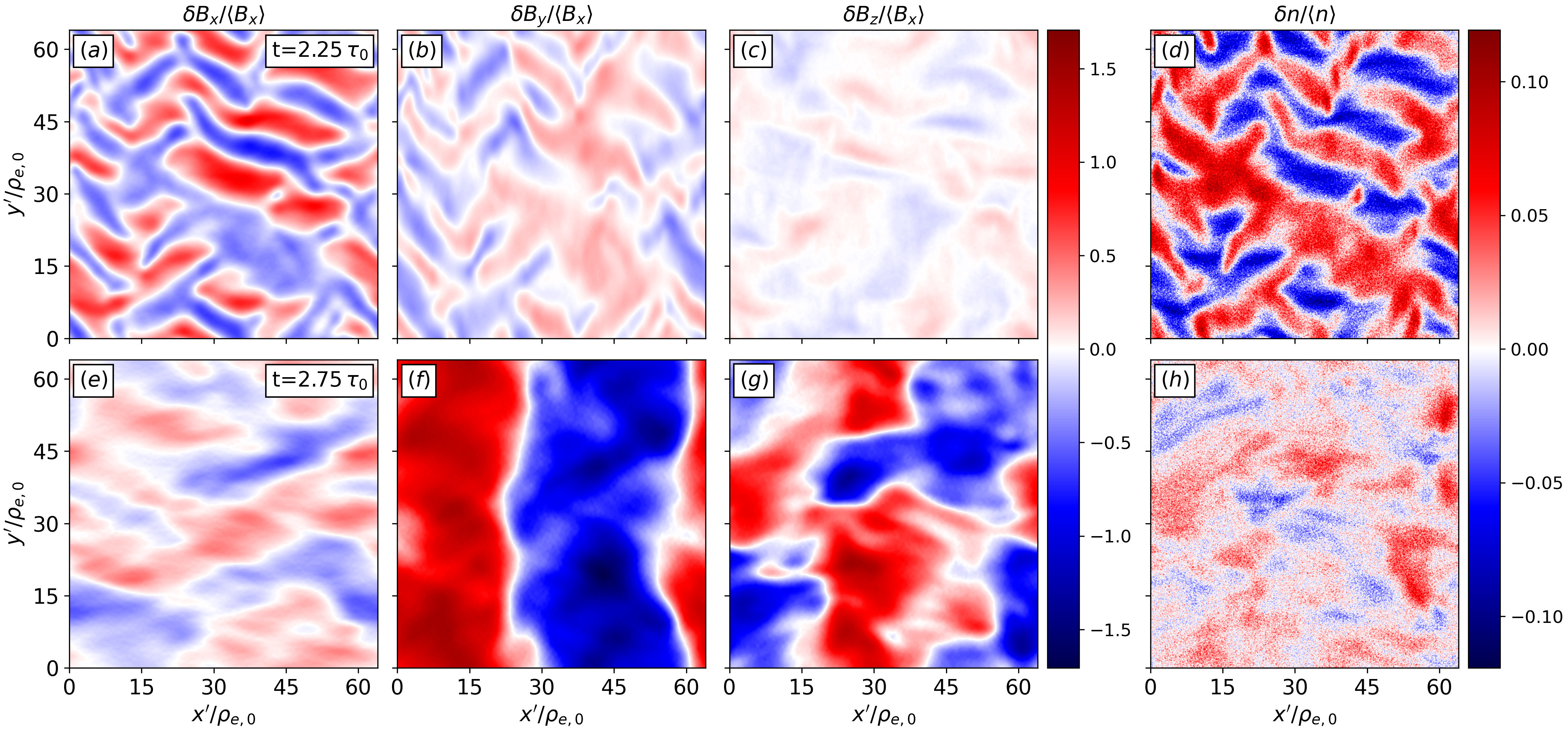}
\caption{Componentwise magnetic-field fluctuations $\delta \bm{B}$ and density fluctuations $\delta n$ at full compression ($t\!=\!2.25\,\tau_0$, top row) and at full expansion ($t\!=\!2.75\,\tau_0$, bottom row), for run $a_0\!=\!0.5$.}
\label{B_n}
\end{figure*}

The spatial structure of instabilities is highlighted in Fig.~\ref{B_n}, showing $\delta \bm{B}$ and density fluctuations $\delta n\!=\!n-\langle n \rangle$ (including all species) for run $a_0\!=\!0.5$, using comoving coordinates $\bm{x}^{\prime}$ (see Supplemental Material \citep{SMPB} for an animation in lab coordinates). Panels (a)-(d) correspond to $t\!=\!2.25\,\tau_0$, when the box is fully compressed. Mirror modes appear as oblique $\delta B_x$ and $\delta B_y$ fluctuations anticorrelated with large $\delta n$ variations, while cyclotron waves produce small $\delta B_z$ perturbations. Panels (e)-(h) correspond to $t\!=\!2.75\,\tau_0$, when the box is fully expanded. Parallel-firehose modes manifest as $\delta B_y$ fluctuations almost uniform in $y^{\prime}$, while oblique-firehose modes perturb $\delta B_z$ and $\delta B_x$, producing small $\delta n$. 

\begin{figure*}[t]
\centering
\includegraphics[width=0.99\linewidth]{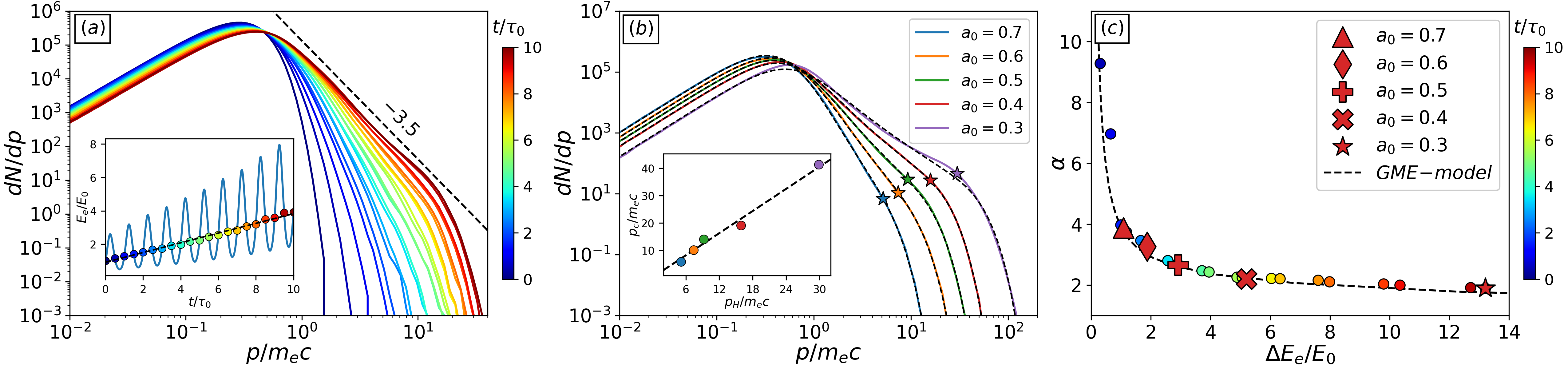}
\caption{(a) Temporal evolution of electron distribution and kinetic energy (inset) of run $a_0\!=\!0.5$. (b) Electron distributions for different $a_0$ at $t\!=\!10\,\tau_0$, with fitted cutoff momenta $p_c$ versus Hillas momenta $p_H$ (inset). (c) Fitted power-law indices $\alpha$ versus injected energy $\Delta E_e/E_0$ for run $a_0\!=\!0.3$ at different times (color-filled dots) and for different $a_0$ at $t\!=\!10\,\tau_0$ (red-filled symbols), compared to our generalized maximum entropy model (dashed-black).}
\label{spectra}
\end{figure*}

The nonadiabatic evolution of $\langle P_\perp \rangle/\langle P_\parallel \rangle$, with instabilities limiting its growth, indicates that unstable modes are isotropizing particle velocity distributions. Hence, energy injected via pulsation is partially channeled through instabilities to particles, eventually accelerating them. Figure~\ref{spectra}(a) shows the temporal evolution of the electron distribution ${\rm d}N/{\rm d}p$ (number of particles per momentum bin, with $p\!=\!|\bm{p}|$ being the momentum \footnote{The corresponding energy distribution is ${\rm d}N/{\rm d}\epsilon\!=\!({\rm d}N/{\rm d}p)({\rm d}p/{\rm d}\epsilon)$, where $\epsilon$ is the kinetic energy.}) of run $a_0\!=\!0.5$, sampled every $\Delta t \!=\! 0.5\,\tau_0$, when the box returns to its initial size. We observe efficient nonthermal particle acceleration, with ${\rm d}N/{\rm d}p$ developing an ultrarelativistic power-law tail, evolving toward ${\rm d}N/{\rm d}p \!\sim\! p^{-3.5}$. The inset shows the temporal evolution of the total electron kinetic energy $E_e(t)$, normalized to $E_0 \!=\! E_e(0)$. Besides oscillating because of compression and expansion, $E_e$ progressively increases after every pulsation cycle, instead of returning to $E_0$, meaning that MP is occurring. A linear fit (dashed-black line) through $E_e$ measured at intervals of $\Delta t \!=\! 0.5\,\tau_0$ (color-filled dots) indicates an energy increase of $\sim\!29\%$ per pulsation cycle (i.e.\ $E_e/E_0\!\simeq\!0.29\,t/\tau_0+1$). Analogous results are found for positrons. We have verified that starting pulsations with expansion rather than compression and varying $\tau_0$ (keeping $\tau_0\!\gg\!\Omega_{e,0}^{-1}$) do not affect particle distributions (see Supplemental Material \citep{SMPB}).

Since free energy available for particle acceleration is determined by the amount of compression, we expect MP to strongly depend on $a_0$. Figure~\ref{spectra}(b) shows electron distributions of runs with different $a_0$, measured at $t\!=\!10\,\tau_0$. As $a_0$ decreases (stronger compression), particle distributions develop harder power-law tails, extending further into ultrarelativistic energies. Following Ref.~\citep{zhdankin2022non}, we model particle distributions using the $\kappa$-distribution 
\begin{equation}
\left( \frac{{\rm d}N}{{\rm d}p} \right)_{\kappa} = C p^2 \left[ 1 + \frac{\epsilon(p)}{\epsilon_b} \right]^{-\alpha-2},
\end{equation}
where $C$ and $\epsilon_b$ are constants, $\epsilon(p)\!=\!(p^2c^2 + m_e^2 c^4)^{1/2} - m_e c^2$, and $\alpha$ is the asymptotic power-law index (i.e.\ $\left( {\rm d}N/{\rm d}p \right)_{\kappa} \!\sim\! p^{-\alpha}$ at ultrarelativistic energies $\epsilon(p)\!\approx\!p c\!\gg\!\epsilon_b$). Since $\left( {\rm d}N/{\rm d}p \right)_{\kappa}$ extends to $p\!\rightarrow\!\infty$, we introduce an energy cutoff $E_c$, as in Ref.~\citep{comisso2024ultra}, of the form 
\begin{equation}
\left( \frac{{\rm d}N}{{\rm d}p} \right)_{\rm fit} = \left( \frac{{\rm d}N}{{\rm d}p} \right)_{\kappa} \!\! \sech\!\left[\frac{\epsilon^2(p)}{\epsilon_c^2}\right],
\label{my_kappa}
\end{equation}
causing a rapid falloff when $\epsilon(p)\!>\!\epsilon_c$. Equation~\eqref{my_kappa} provides an excellent fit for electron distributions, as seen in Fig.~\ref{spectra}(b), showing fitted $\left( {\rm d}N/{\rm d}p \right)_{\rm fit}$ (dashed-black curves, see Supplemental Material \citep{SMPB} for fit parameters) plotted over their corresponding ${\rm d}N/{\rm d}p$ from simulations. Physically, the high-energy cutoff is determined by the Hillas limit \citep{hillas1984origin}, implying that acceleration becomes inefficient when a particle's gyroradius reaches the size of the accelerator $\lambda$, resulting in a maximum momentum $p_H\!=\!e \langle B \rangle \lambda/c$ (where $\langle B \rangle$ is the mean magnetic-field amplitude). Since acceleration by MP is mediated by instabilities, $\lambda$ corresponds to the magnetic-field integral-scale
\begin{equation}
\lambda_B = 2\pi \left( \frac{\bigintsss \sqrt{k_x^2 + k_y^2}\, P_B\, {\rm d}k_x {\rm d}k_y}{\bigintsss P_B\, {\rm d}k_x {\rm d}k_y} \right)^{-1},   
\end{equation}
where $P_B(k_x,\,k_y)$ is the power spectrum of magnetic fluctuations with wavevectors $\bm{k}\!=\!(k_x,\,k_y)$. To account for $\langle B \rangle$ and $\lambda_B$ variations induced by pulsation, we define the pulsating box Hillas momentum as $p_H \!=\! {\rm max}(e \langle B \rangle \lambda_B/c)$, corresponding to the largest $e \langle B \rangle \lambda_B/c$ measured over time (reached after $\sim\!5\,\tau_0$ for all runs, see Supplemental Material \citep{SMPB}). Figure~\ref{spectra}(b) shows locations of $p_H$ (color-filled stars) on their corresponding distribution ${\rm d}N/{\rm d}p$ for different $a_0$, appropriately indicating where power-law tails start declining. The inset shows that fitted cutoff momenta $p_c\!=\![(\epsilon_c + m_e c^2)^2 - m_e^2 c^4]^{1/2}/c$ and $p_H$ are proportional, with the dashed-black line indicating $p_c\!=\!1.35\, p_H$. 

The development of $\kappa$-distributions driven by MP suggests a connection to the GME principle formulated in Ref.~\citep{zhdankin2022non}, extending the thermodynamic maximum entropy principle to nonequilibrium collisionless systems. The GME model predicts that particles relax toward $\kappa$-distributions to maximize Casimir momenta \citep{zhdankin2022generalized}, representing extensions of the Boltzmann--Gibbs entropy \citep{zhdankin2023dimensional}. Using the GME framework, we derive a model describing how the power-law index of the $\kappa$-distribution evolves with the injected energy. We consider the ultrarelativistic $\kappa$-distribution \footnote{We choose an ultrarelativistic distribution because it is appropriate to describe the tail of electron distributions from simulations, extending into the $p\!\gg\!m_e c$ range, and because it allows for the model to be derived analytically in closed form.}
\begin{equation}
\frac{{\rm d}N}{{\rm d}p} = 4\pi p^2 f_0 \left( 1 + \frac{p}{p_b} \right)^{-\alpha-2},    
\end{equation}
with hard cutoff at $p_c$ such that ${\rm d}N/{\rm d}p\!=\!0$ at $p\!>\!p_c$. Here, $f_0$ is the peak of the distribution in 3D momentum space, at $p\!=\!0$, and $p_b$ is the break momentum, governing the transition between core and tail. Since MP is typically described as a diffusion process with diffusion coefficient $D\!\sim\!p^2$ \citep{lichko2017magnetic,lichko2020magnetic,malkov2026magnetic}, we expect particles with $p\!\simeq\!0$ to be weakly affected by MP. Thus, we assume that as energy is injected into the system, $f_0$ remains approximately constant in time while $p_b$ and $\alpha$ evolve, subject to constraints on the number of particles $N$ and total kinetic energy $E$. Assuming $p_c\!\gg\!p_b$ and $\alpha\!>\!1$ \footnote{$\alpha\!\leqslant\!1$ would cause $N$ to diverge in large-scale systems, where $p_c\!\rightarrow\!\infty$.}, we calculate
\begin{equation}
N = \int_0^{p_c} \frac{{\rm d}N}{{\rm d}p} {\rm d}p \simeq \frac{8\pi f_0 p_b^3}{\alpha^3-\alpha},
\label{NGME}
\end{equation}
and
\begin{equation}
\begin{aligned}
E & = \int_0^{p_c} p c \frac{{\rm d}N}{{\rm d}p} {\rm d}p 
\\ 
&\simeq \frac{24\pi f_0 p_b^4 c}{(\alpha^3-\alpha)(\alpha-2)} \left[ 1 - \frac{\alpha^3 - \alpha}{6} \left(\frac{p_b}{p_c} \right)^{\alpha-2} \right].
\end{aligned}    
\label{EGME}
\end{equation}
We assume that ${\rm d}N/{\rm d}p$ is initially Maxwell--J{\"u}ttner, so that energy must satisfy the initial condition $E\!=\!E_0$ at $\alpha\!\rightarrow\!\infty$, giving 
\begin{equation}
E_0 \simeq \frac{3c N^{4/3}}{2 \pi^{1/3} f_0^{1/3}} \quad\implies\quad f_0 \simeq \frac{27 c^3 N^4}{8 \pi E_0^3}. 
\label{fGME}
\end{equation}
Combining Eqs.~\eqref{NGME}-\eqref{fGME}, we obtain
\begin{equation}
\begin{aligned}
\frac{\Delta E}{E_0} = & \frac{(\alpha^3-\alpha)^{1/3}}{\alpha-2} - 1 \\
& -\frac{(\alpha^3-\alpha)^{(\alpha+2)/3}}{6(\alpha-2)}\left( \frac{E_0}{3 E_c} \right)^{\alpha-2},
\end{aligned}
\label{dEGME}
\end{equation}
describing the universal evolution of $\alpha$ with $\Delta E\!=\!E-E_0$, and $E_c\!=\!N p_c c$ denoting the maximum kinetic energy achievable.

Figure~\ref{spectra}(c) shows power-law indices $\alpha$ obtained by fitting electron distributions from simulations with Eq.~\eqref{my_kappa}, versus $\Delta E_e/E_0\!=\!(E_e-E_0)/E_0$. Simulations are compared to Eq.~\eqref{dEGME}, with $p_c$ equal to $p_H$ of run $a_0\!=\!0.3$ (the curve is insensitive to $p_c$ for $\Delta E_e/E_0\!<\!5$). The temporal evolution of $\alpha$ of run $a_0\!=\!0.3$ (color-filled dots) is in excellent agreement with our GME model. Furthermore, $\alpha$ of runs with different $a_0$, measured at $t\!=\!10\,\tau_0$, also match Eq.~\eqref{dEGME} very well. In addition to $a_0$, the $\Delta E$ gained per pulsation cycle also depends on how quickly instabilities isotropize the pressure, with faster isotropization rates implying a larger energy gain. However, even for slow isotropization, energy would still be injected and the GME model would apply.

{\it Conclusions---}We have investigated, for the first time, particle acceleration in high-$\beta$ plasmas undergoing compression--expansion (``pulsation'') cycles, using fully kinetic simulations implementing a new ``pulsating box'' setup. Pulsations produce pressure-anisotropy variations, triggering kinetic instabilities that channel the injected energy to particles, efficiently accelerating them via MP. Our results confirm that MP is a viable mechanism for particle acceleration, capable of producing nonthermal distributions with high-energy power-law tails. Hence, MP could be an important source of high-energy particles in high-$\beta$ space and astrophysical plasmas such as planetary magnetosheaths, accretion flows, and the intracluster medium, where $10\!\lesssim\!\beta\!\lesssim\!100$ and reconnection and turbulence are typically inefficient particle accelerators. We expect our results (including agreement with the GME model) to hold over a broad $\beta\!\gg\!1$ range, where even small departures from $\langle P_\perp \rangle/\langle P_\parallel \rangle \!=\!1$ produce instabilities that could drive MP.

Our numerical results are in excellent agreement with a GME model that we have derived, allowing for calculating particle distributions by knowing only the relative amount of energy injected in the system $\Delta E/E_0$ and the maximum achievable energy $E_c/E_0$. These parameters can be potentially measured or estimated for many systems, making our model a powerful tool capable of predicting distributions of high-energy particles in space and astrophysical plasmas. Furthermore, our model can be scaled to the macroscopic size of space and astrophysical sources by adjusting $E_c$, related to the Hillas limit imposed by the system size. Specifically, in large systems, our model becomes insensitive to $E_c$, and Eq.~\eqref{dEGME} simplifies to
\begin{equation}
\frac{\Delta E}{E_0} \simeq \frac{(\alpha^3-\alpha)^{1/3}}{\alpha-2} - 1,
\end{equation}
predicting an asymptotic power-law tail with index $\alpha\!=\!2$ when $\Delta E/E_0\!\gg\!1$. Our GME model goes beyond previous MP models where the plasma is typically assumed to be mostly adiabatic and only weakly perturbed by instabilities.

We note that the derivation of our model is purely based on statistical mechanics arguments and is agnostic of specific plasma processes accelerating particles. Consequently, we expect GME models to be applicable to a wide variety of plasma phenomena \citep{zhdankin2022non,zhdankin2022generalized}. Besides GME, other models have been developed to extend the maximum entropy principle to nonequilibrium systems, predicting $\kappa$-distributions as the relaxed state of plasmas \citep{tsallis1988possible,livadiotis2013understanding,livadiotis2018thermodynamic,livadiotis2021thermodynamic}; however, our model provides novel predictions connecting $\kappa$-distribution parameters to physical parameters.

Our GME model is applicable to 3D geometries and multispecies plasmas, since it does not require any assumption regarding the spatial dimensionality and the plasma composition. In future studies, we plan to investigate MP driven by pulsation in 3D domains. A 3D geometry would allow for more oblique mirror and firehose modes to develop and interact (facilitating turbulence), potentially making particle acceleration by MP faster than in 2D setups discussed in this work. We also plan to extend our study to ion-electron plasmas, investigating how MP accelerates particle species with different mass and temperature ratios (with these parameters governing instability thresholds \citep{gary1993theory} and the energy partition among different species \citep{zhdankin2019electron,arro2022spectral}). We expect that the GME model would still predict the power-law indices of ion and electron distributions separately, but not the ion-to-electron energy partition, typically determined by plasma processes. Finally, we plan to extend our work by including inhomogeneities, turbulence and collisions, that could further enhance MP, even in low-$\beta$ plasmas \citep{lichko2017magnetic,malkov2026magnetic}.

\begin{acknowledgments}

{\it Acknowledgments---}GA is supported by the NASA grant No. 80NSSC26K0594. Opinions, findings, and conclusions expressed in this work are those of the authors and do not necessarily reflect the views of NASA. The authors acknowledge support from the National Science Foundation under NSF grant PHY-2409316, and the Department of Energy under the grant DE-SC0026099. This work used Stampede3 at the Texas Advanced Computer Center (TACC) through allocation PHY160032 from the Advanced Cyberinfrastructure Coordination Ecosystem: Services \& Support (ACCESS) program, which is supported by U.S. National Science Foundation grants \#2138259, \#2138286, \#2138307, \#2137603, and \#2138296. FB acknowledges support from the FED-tWIN programme (profile Prf-2020-004, project ``ENERGY'') issued by BELSPO, and from the FWO Junior Research Project G020224N granted by the Research Foundation -- Flanders (FWO).

\end{acknowledgments}

\appendix 

\section{\large Supplemental Material}

\section{Temporal evolution of magnetic-field fluctuations}

The file ``PulsatingBox.gif'' shows an animation of the componentwise temporal evolution of magnetic-field fluctuations $\delta \bm{B}\!=\!\bm{B}-\langle\bm{B}\rangle$ (where $\langle\cdot\rangle$ denotes volume averaging) in lab coordinates $\bm{x}$, for our fiducial run $a_0\!=\!0.5$, presented in the main manuscript. 

\section{Pulsations starting with expansion}
The results presented in the main manuscript consider pulsation cycles starting with compression, meaning that the coordinate transformation describing pulsations
\begin{equation}
\bm{x} = \textbf{L}\,\bm{x}^{\prime} =
\begin{pmatrix}
1 & 0 & 0 \\
0 & a & 0 \\
0 & 0 & a
\end{pmatrix}
\bm{x}^{\prime},
\quad\quad a = a_0^{\sin(2\pi\,t/\tau_0)},
\label{matrix2}
\end{equation}
has $a_0\!<\!1$. On the other hand, using $a_0\!>\!1$ implies that pulsation cycles start with expansion. To test if results presented in the main manuscript depend on whether pulsation cycles start with compression or expansion, we compare two runs, one with $a_0\!=\!2$ (i.e.\ starting with expansion) and our fiducial run with $a_0\!=\!0.5$ (i.e.\ starting with compression), keeping all the other parameters the same. In both runs, the maximally compressed state corresponds to compressing the box size in the $y$-direction to $ L_y/2$ (where $L_y$ is the initial box size), while the maximally expanded state corresponds to expanding the box size to $2L_y$. 

\begin{figure}[t]
\centering
\includegraphics[width=0.99\linewidth]{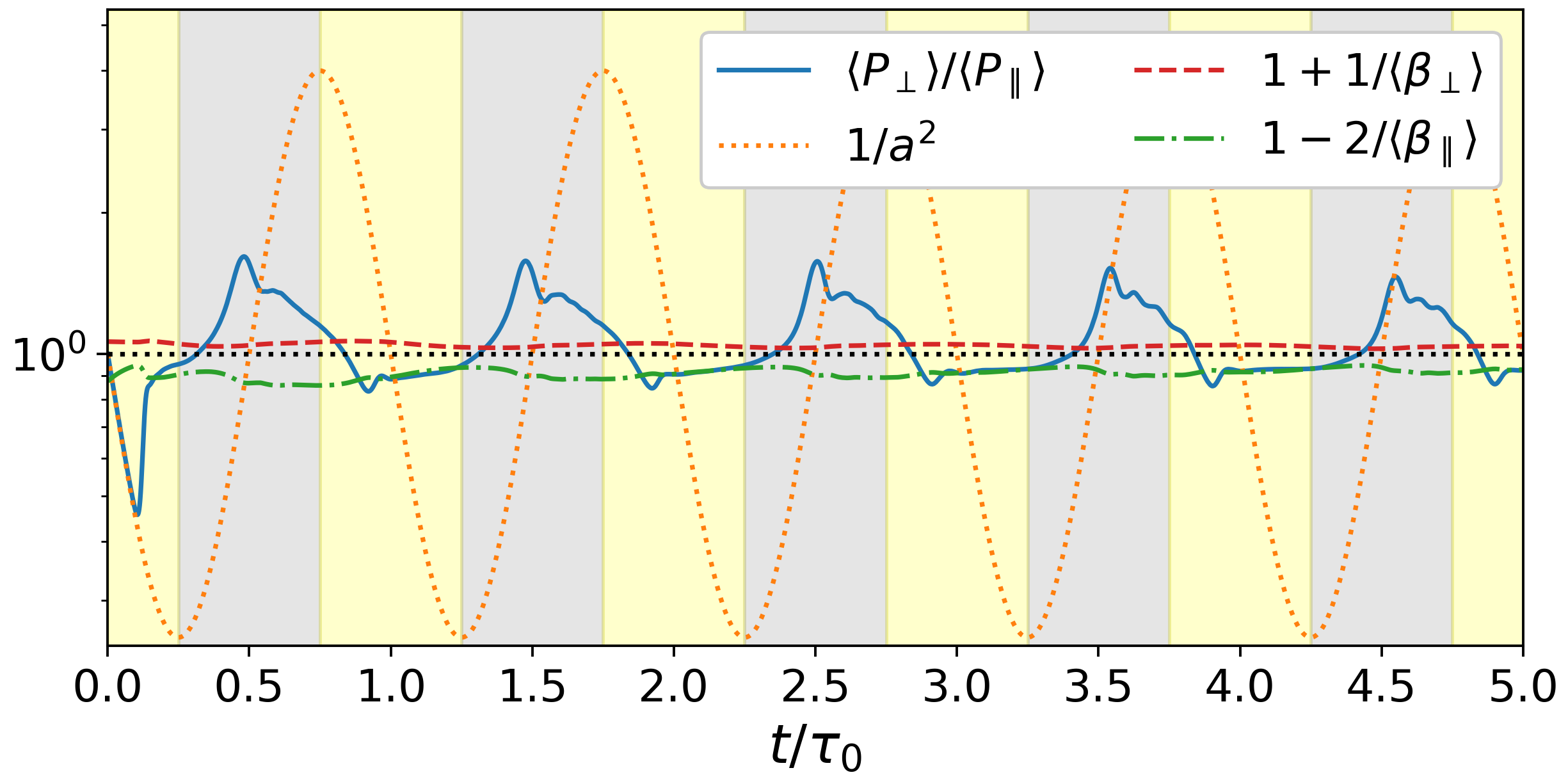}
\caption{Temporal evolution of run $a_0\!=\!2.0$. Pressure-anisotropy (solid-blue), adiabatic scaling (dotted-orange), mirror (dashed-red) and firehose (dash-dotted-green) instability thresholds, $\langle P_\perp \rangle/\langle P_\parallel \rangle \!=\! 1$ (dotted black). Gray(yellow)-shaded areas highlight compression (expansion) cycles.}
\label{anisotropy2}
\end{figure}

Figure~\ref{anisotropy2} shows the temporal evolution of the pressure-anisotropy $\langle P_\perp \rangle/\langle P_\parallel \rangle$ (solid-blue) of run $a_0\!=\!2$, compared to the subrelativistic adiabatic scaling $1/a^2$ (dotted-orange), and to mirror and firehose-instability thresholds, $1+1/\langle \beta_\perp \rangle$ (dashed-red) and $1-2/\langle \beta_\parallel \rangle$ (dash-dotted-green), respectively. We see that, after a brief initial adiabatic expansion, $\langle P_\perp \rangle/\langle P_\parallel \rangle$ oscillates nonadiabatically within a narrower range than $1/a^2$, limited by the periodic excitation of instabilities. Specifically, $\langle P_\perp \rangle/\langle P_\parallel \rangle$ exceeds the mirror-instability threshold when $\langle P_\perp \rangle/\langle P_\parallel \rangle \!>\! 1+1/\langle \beta_\perp \rangle$, during compression (gray-shaded areas), and the firehose instability threshold when $\langle P_\perp \rangle/\langle P_\parallel \rangle \!<\! 1-2/\langle \beta_\parallel \rangle$, during expansion (yellow-shaded areas). Comparing the temporal evolution of $\langle P_\perp \rangle/\langle P_\parallel \rangle$ of run $a_0\!=\!2$ with that of run $a_0\!=\!0.5$ (in the main manuscript), we find that after $\sim\!0.25\,\tau_0$, pressure-anisotropy starts following the same periodic evolution, independently of how pulsation cycles start.

\begin{figure}[h]
\centering
\includegraphics[width=0.99\linewidth]{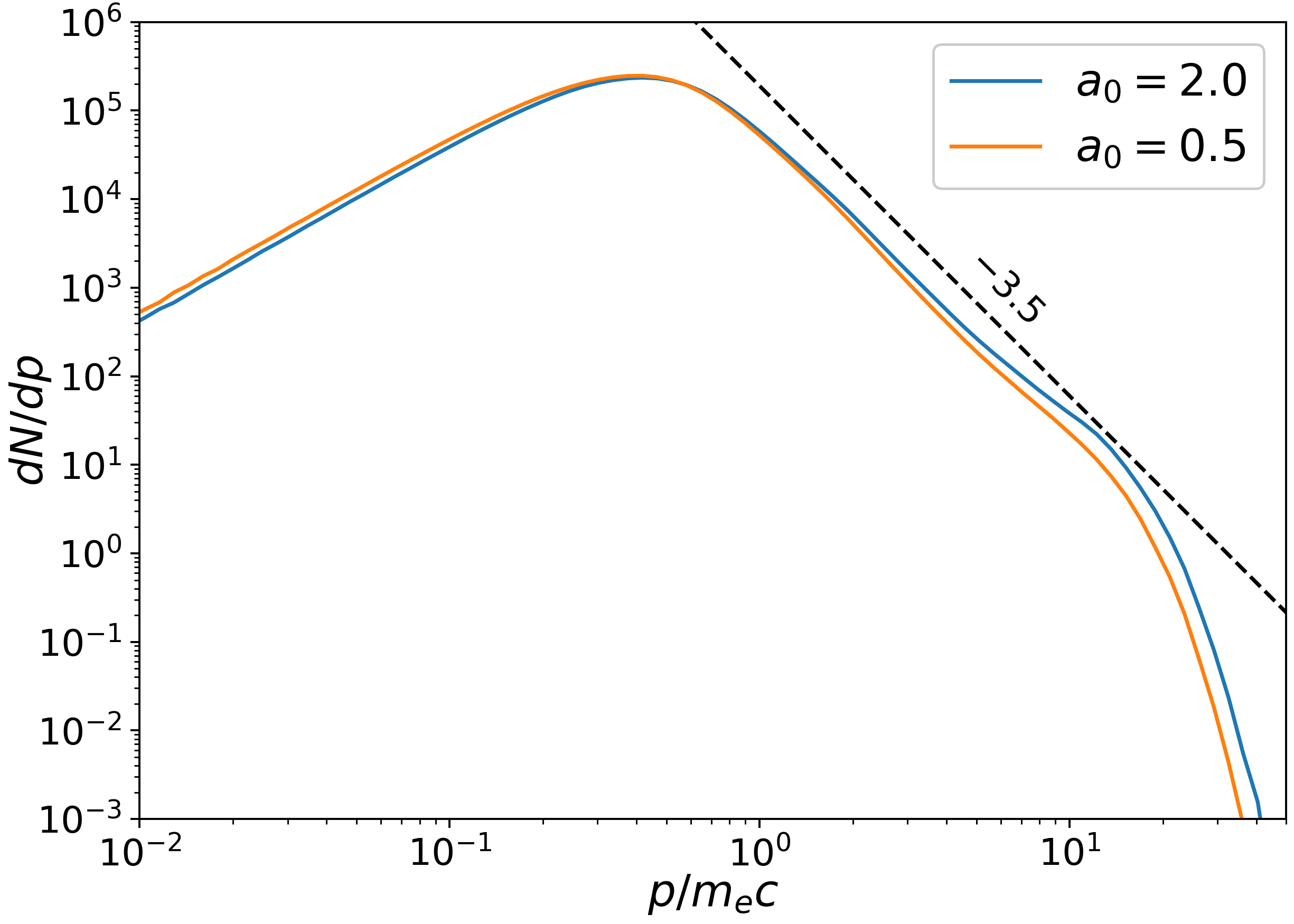}
\caption{Electron distributions of runs $a_0\!=\!2.0$ (solid blue) and $a_0\!=\!0.5$ (solid orange), measured at $t\!=\!10\,\tau_0$}
\label{spectra2}
\end{figure}

In Fig.~\ref{spectra2}, we compare electron distributions ${\rm d}N/{\rm d}p$ (as functions of momentum $p\!=\!|\textbf{p}|$) of runs $a_0\!=\!2$ (solid blue) and $a_0\!=\!0.5$ (solid orange), measured at $t\!=\!10\,\tau_0$. We find that both distributions are essentially equivalent, meaning that particle acceleration is independent of whether pulsation cycles start with expansion or compression.

\section{Simulations with different pulsation periods}

To test whether results presented in the main manuscript depend on the pulsation period $\tau_0$, we compare runs with $a_0\!=\!0.5$ and different $\tau_0$, keeping all the other parameters the same as those used for our fiducial run. 

\begin{figure}[h]
\centering
\includegraphics[width=0.99\linewidth]{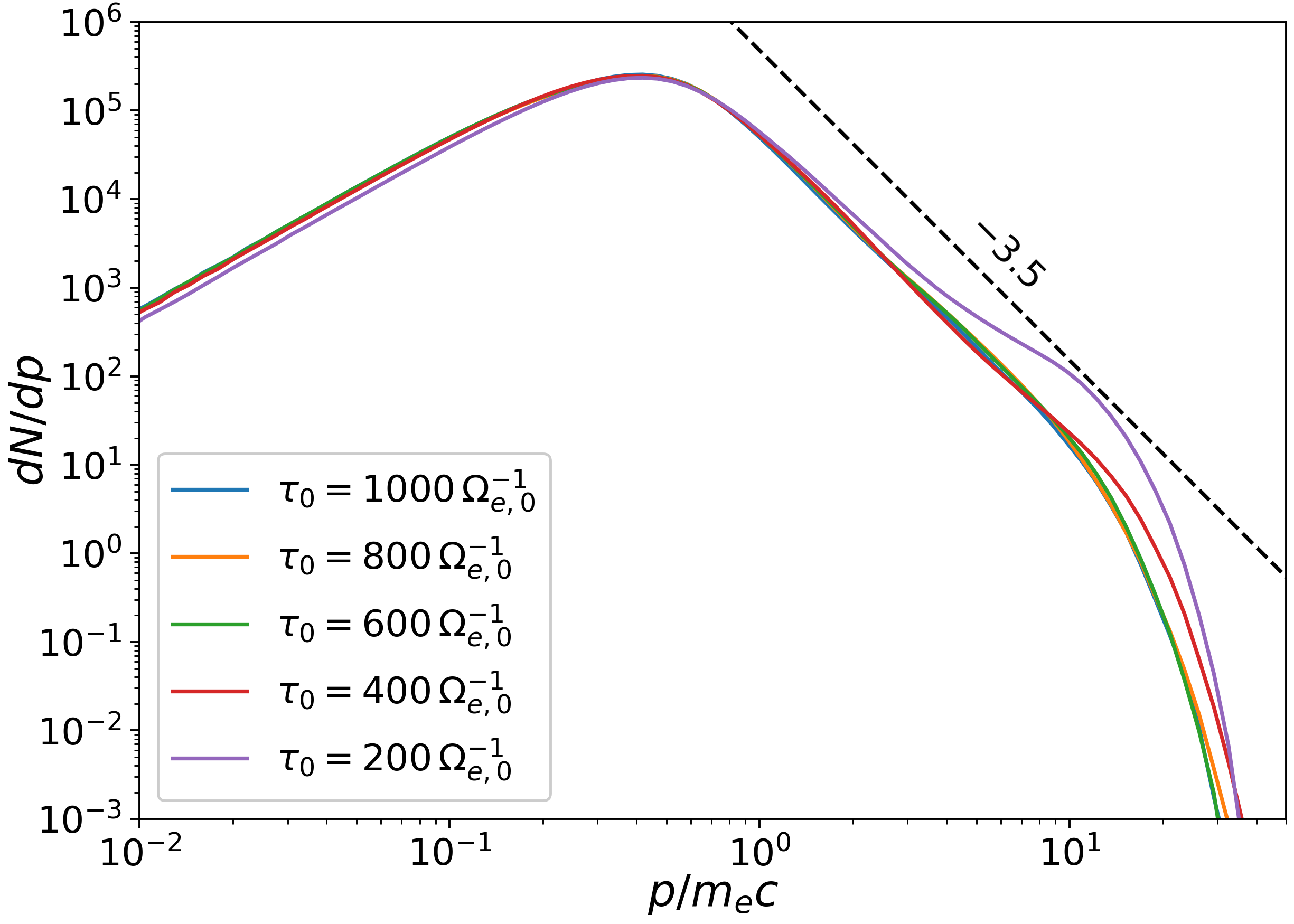}
\caption{Electron distributions of runs with different $\tau_0$, measured at $t\!=\!10\,\tau_0$}
\label{spectra_tau0}
\end{figure}

Figure~\ref{spectra_tau0} shows electron distributions ${\rm d}N/{\rm d}p$ of runs with different $\tau_0$ (with the latter in units of the initial electron gyroperiod $\Omega_{e,0}^{-1}$), measured at $t\!=\!10\,\tau_0$. We find that distributions are essentially independent of $\tau_0$, except for run $\tau_0\!=\!200\,\Omega_{e,0}^{-1}$, showing a pileup of particles at high energies, where the power-law tail starts declining. The behavior of run $\tau_0\!=\!200\,\Omega_{e,0}^{-1}$ is most likely caused by the fact that the expanding/compressing box model we are using is valid when compression and expansion timescales (i.e.\ $\tau_0$ in our setup) are much larger than the particle gyroperiod. When $\tau_0$ approaches $\Omega_{e,0}^{-1}$, the model must be modified by introducing extra fictitious forces proportional to $\ddot{a}\!=\!{\rm d}^2a/{\rm d}t^2$, accounting for the box acceleration during compression and expansion, as discussed in Ref.~\citep{sironi2015electron}.

\section{Particle distribution fit parameters}

Table~\ref{table} reports fit parameters (obtained via a least-squares fit) for particle distributions of simulations with different $a_0$ (at $t\!=\!10\,\tau_0$), shown in Fig.~3(b) in the main manuscript. Here, the break energy $\epsilon_b$ and cutoff energy $\epsilon_c$ are normalized to the electron rest mass energy $m_e c^2$.

\begin{table}[t]
\caption{Fit parameters for particle distributions shown in Fig.~3(b) in the main manuscript.}
\label{table}
\begin{ruledtabular}
\begin{tabular}{lcdr}
\textrm{$a_0$}&
\textrm{$\epsilon_b/(m_e c^2)$}&
\textrm{$\alpha$}&
\textrm{$\epsilon_c/(m_e c^2)$}\\
\colrule
0.7 & 0.24 & 3.9 & 4.9\\
0.6 & 0.25 & 3.3 & 9.1\\
0.5 & 0.25 & 2.6 & 13.0\\
0.4 & 0.26 & 2.2 & 18.1\\
0.3 & 0.32 & 1.9 & 40.4\\
\end{tabular}
\end{ruledtabular}
\end{table}

\section{Temporal evolution of the Hillas momentum}

\begin{figure}[h]
\centering
\includegraphics[width=0.99\linewidth]{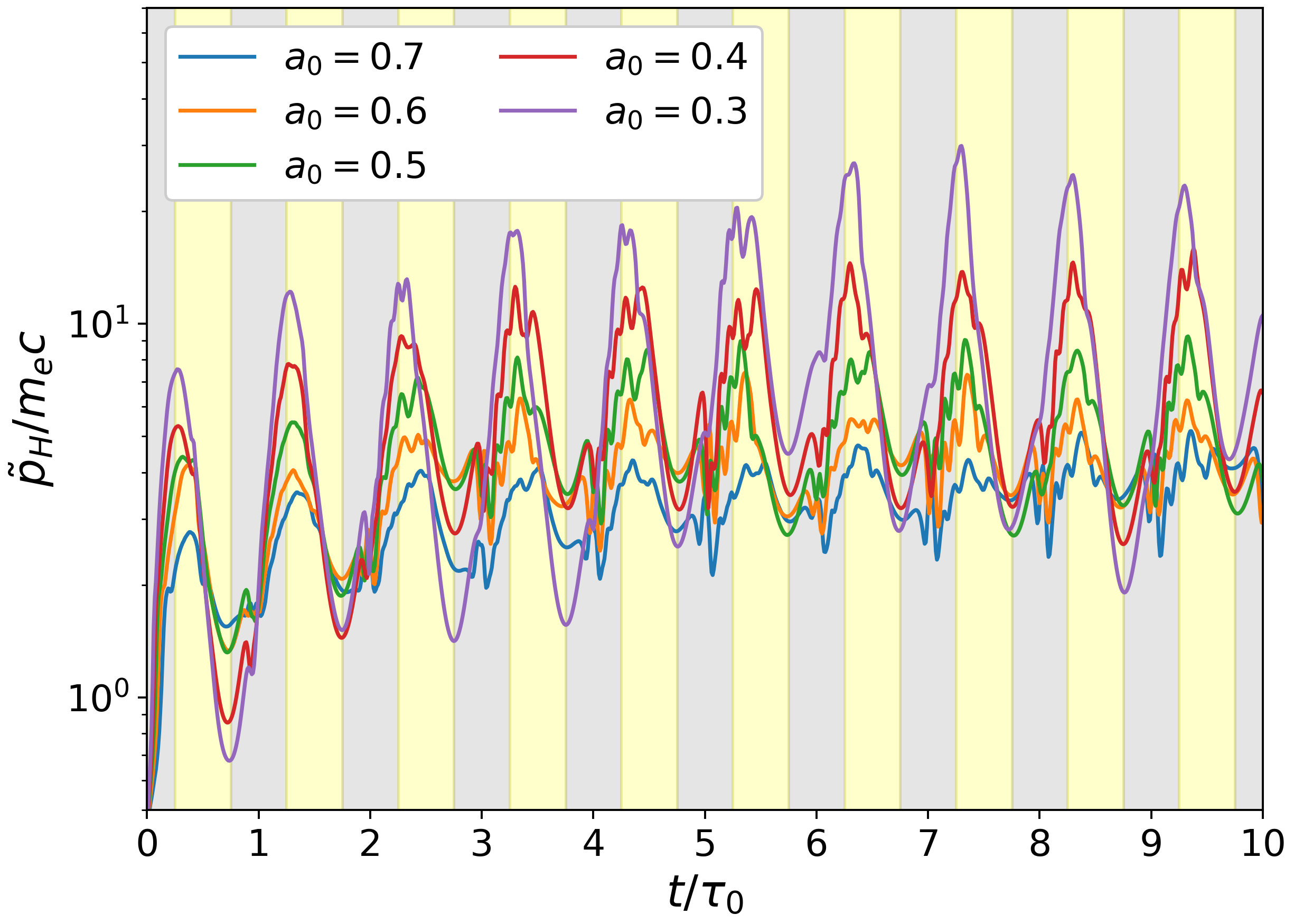}
\caption{Temporal evolution of $\tilde{p}_H$ of runs with different $a_0$.}
\label{pmax}
\end{figure}

Figure~\ref{pmax} shows the temporal evolution of the Hillas momentum $\tilde{p}_H\!=\!e\langle B \rangle \lambda_B/c$ (where $\lambda_B$ is the magnetic-field integral-scale) of runs with different $a_0$. We see that $\tilde{p}_H$ oscillates over time, peaking precisely between the end of compression phases (gray-shaded areas) and the very beginning of expansion phases (yellow-shaded areas). As $a_0$ decreases (stronger compression), $\tilde{p}_H$ achieves larger values, meaning that higher levels of compression result in particle acceleration to higher energies. We note that $\tilde{p}_H$ reaches its maximum after $\sim\!5\,\tau_0$ for all runs. This implies that particles can reach the maximum energy allowed by the Hillas limit, i.e.\ $p_H\!=\!{\rm max}(\tilde{p}_H)$, way before $10$ full pulsation cycles. Note that, even if particles can access their maximum energy after $\sim\!5\,\tau_0$, it still takes time for magnetic pumping to accelerate a significant amount of particles and to populate the high-energy power-law tail of the distribution. 

\begin{thebibliography}{57}%
\makeatletter
\providecommand \@ifxundefined [1]{%
 \@ifx{#1\undefined}
}%
\providecommand \@ifnum [1]{%
 \ifnum #1\expandafter \@firstoftwo
 \else \expandafter \@secondoftwo
 \fi
}%
\providecommand \@ifx [1]{%
 \ifx #1\expandafter \@firstoftwo
 \else \expandafter \@secondoftwo
 \fi
}%
\providecommand \natexlab [1]{#1}%
\providecommand \enquote  [1]{``#1''}%
\providecommand \bibnamefont  [1]{#1}%
\providecommand \bibfnamefont [1]{#1}%
\providecommand \citenamefont [1]{#1}%
\providecommand \href@noop [0]{\@secondoftwo}%
\providecommand \href [0]{\begingroup \@sanitize@url \@href}%
\providecommand \@href[1]{\@@startlink{#1}\@@href}%
\providecommand \@@href[1]{\endgroup#1\@@endlink}%
\providecommand \@sanitize@url [0]{\catcode `\\12\catcode `\$12\catcode `\&12\catcode `\#12\catcode `\^12\catcode `\_12\catcode `\%12\relax}%
\providecommand \@@startlink[1]{}%
\providecommand \@@endlink[0]{}%
\providecommand \url  [0]{\begingroup\@sanitize@url \@url }%
\providecommand \@url [1]{\endgroup\@href {#1}{\urlprefix }}%
\providecommand \urlprefix  [0]{URL }%
\providecommand \Eprint [0]{\href }%
\providecommand \doibase [0]{https://doi.org/}%
\providecommand \selectlanguage [0]{\@gobble}%
\providecommand \bibinfo  [0]{\@secondoftwo}%
\providecommand \bibfield  [0]{\@secondoftwo}%
\providecommand \translation [1]{[#1]}%
\providecommand \BibitemOpen [0]{}%
\providecommand \bibitemStop [0]{}%
\providecommand \bibitemNoStop [0]{.\EOS\space}%
\providecommand \EOS [0]{\spacefactor3000\relax}%
\providecommand \BibitemShut  [1]{\csname bibitem#1\endcsname}%
\let\auto@bib@innerbib\@empty
\bibitem [{\citenamefont {McConnell}\ \emph {et~al.}(2002)\citenamefont {McConnell}, \citenamefont {Zdziarski}, \citenamefont {Bennett}, \citenamefont {Bloemen}, \citenamefont {Collmar}, \citenamefont {Hermsen}, \citenamefont {Kuiper}, \citenamefont {Paciesas}, \citenamefont {Phlips}, \citenamefont {Poutanen} \emph {et~al.}}]{mcconnell2002soft}%
  \BibitemOpen
  \bibfield  {author} {\bibinfo {author} {\bibfnamefont {M.~L.}\ \bibnamefont {McConnell}}, \bibinfo {author} {\bibfnamefont {A.}~\bibnamefont {Zdziarski}}, \bibinfo {author} {\bibfnamefont {K.}~\bibnamefont {Bennett}}, \bibinfo {author} {\bibfnamefont {H.}~\bibnamefont {Bloemen}}, \bibinfo {author} {\bibfnamefont {W.}~\bibnamefont {Collmar}}, \bibinfo {author} {\bibfnamefont {W.}~\bibnamefont {Hermsen}}, \bibinfo {author} {\bibfnamefont {L.}~\bibnamefont {Kuiper}}, \bibinfo {author} {\bibfnamefont {W.}~\bibnamefont {Paciesas}}, \bibinfo {author} {\bibfnamefont {B.}~\bibnamefont {Phlips}}, \bibinfo {author} {\bibfnamefont {J.}~\bibnamefont {Poutanen}}, \emph {et~al.},\ }\href@noop {} {\bibfield  {journal} {\bibinfo  {journal} {The Astrophysical Journal}\ }\textbf {\bibinfo {volume} {572}},\ \bibinfo {pages} {984} (\bibinfo {year} {2002})}\BibitemShut {NoStop}%
\bibitem [{\citenamefont {Fisk}\ and\ \citenamefont {Gloeckler}(2012)}]{fisk2012particle}%
  \BibitemOpen
  \bibfield  {author} {\bibinfo {author} {\bibfnamefont {L.}~\bibnamefont {Fisk}}\ and\ \bibinfo {author} {\bibfnamefont {G.}~\bibnamefont {Gloeckler}},\ }\href@noop {} {\bibfield  {journal} {\bibinfo  {journal} {Space science reviews}\ }\textbf {\bibinfo {volume} {173}},\ \bibinfo {pages} {433} (\bibinfo {year} {2012})}\BibitemShut {NoStop}%
\bibitem [{\citenamefont {Oka}\ \emph {et~al.}(2018)\citenamefont {Oka}, \citenamefont {Birn}, \citenamefont {Battaglia}, \citenamefont {Chaston}, \citenamefont {Hatch}, \citenamefont {Livadiotis}, \citenamefont {Imada}, \citenamefont {Miyoshi}, \citenamefont {Kuhar}, \citenamefont {Effenberger} \emph {et~al.}}]{oka2018electron}%
  \BibitemOpen
  \bibfield  {author} {\bibinfo {author} {\bibfnamefont {M.}~\bibnamefont {Oka}}, \bibinfo {author} {\bibfnamefont {J.}~\bibnamefont {Birn}}, \bibinfo {author} {\bibfnamefont {M.}~\bibnamefont {Battaglia}}, \bibinfo {author} {\bibfnamefont {C.}~\bibnamefont {Chaston}}, \bibinfo {author} {\bibfnamefont {S.}~\bibnamefont {Hatch}}, \bibinfo {author} {\bibfnamefont {G.}~\bibnamefont {Livadiotis}}, \bibinfo {author} {\bibfnamefont {S.}~\bibnamefont {Imada}}, \bibinfo {author} {\bibfnamefont {Y.}~\bibnamefont {Miyoshi}}, \bibinfo {author} {\bibfnamefont {M.}~\bibnamefont {Kuhar}}, \bibinfo {author} {\bibfnamefont {F.}~\bibnamefont {Effenberger}}, \emph {et~al.},\ }\href@noop {} {\bibfield  {journal} {\bibinfo  {journal} {Space Science Reviews}\ }\textbf {\bibinfo {volume} {214}},\ \bibinfo {pages} {82} (\bibinfo {year} {2018})}\BibitemShut {NoStop}%
\bibitem [{\citenamefont {Salem}\ \emph {et~al.}(2023)\citenamefont {Salem}, \citenamefont {Pulupa}, \citenamefont {Bale},\ and\ \citenamefont {Verscharen}}]{salem2023precision}%
  \BibitemOpen
  \bibfield  {author} {\bibinfo {author} {\bibfnamefont {C.~S.}\ \bibnamefont {Salem}}, \bibinfo {author} {\bibfnamefont {M.}~\bibnamefont {Pulupa}}, \bibinfo {author} {\bibfnamefont {S.~D.}\ \bibnamefont {Bale}},\ and\ \bibinfo {author} {\bibfnamefont {D.}~\bibnamefont {Verscharen}},\ }\href@noop {} {\bibfield  {journal} {\bibinfo  {journal} {Astronomy \& Astrophysics}\ }\textbf {\bibinfo {volume} {675}},\ \bibinfo {pages} {A162} (\bibinfo {year} {2023})}\BibitemShut {NoStop}%
\bibitem [{\citenamefont {Cranmer}\ and\ \citenamefont {Winebarger}(2019)}]{cranmer2019properties}%
  \BibitemOpen
  \bibfield  {author} {\bibinfo {author} {\bibfnamefont {S.~R.}\ \bibnamefont {Cranmer}}\ and\ \bibinfo {author} {\bibfnamefont {A.~R.}\ \bibnamefont {Winebarger}},\ }\href@noop {} {\bibfield  {journal} {\bibinfo  {journal} {Annual Review of Astronomy and Astrophysics}\ }\textbf {\bibinfo {volume} {57}},\ \bibinfo {pages} {157} (\bibinfo {year} {2019})}\BibitemShut {NoStop}%
\bibitem [{\citenamefont {Kasper}\ \emph {et~al.}(2021)\citenamefont {Kasper}, \citenamefont {Klein}, \citenamefont {Lichko}, \citenamefont {Huang}, \citenamefont {Chen}, \citenamefont {Badman}, \citenamefont {Bonnell}, \citenamefont {Whittlesey}, \citenamefont {Livi}, \citenamefont {Larson} \emph {et~al.}}]{kasper2021parker}%
  \BibitemOpen
  \bibfield  {author} {\bibinfo {author} {\bibfnamefont {J.}~\bibnamefont {Kasper}}, \bibinfo {author} {\bibfnamefont {K.}~\bibnamefont {Klein}}, \bibinfo {author} {\bibfnamefont {E.}~\bibnamefont {Lichko}}, \bibinfo {author} {\bibfnamefont {J.}~\bibnamefont {Huang}}, \bibinfo {author} {\bibfnamefont {C.}~\bibnamefont {Chen}}, \bibinfo {author} {\bibfnamefont {S.}~\bibnamefont {Badman}}, \bibinfo {author} {\bibfnamefont {J.}~\bibnamefont {Bonnell}}, \bibinfo {author} {\bibfnamefont {P.}~\bibnamefont {Whittlesey}}, \bibinfo {author} {\bibfnamefont {R.}~\bibnamefont {Livi}}, \bibinfo {author} {\bibfnamefont {D.}~\bibnamefont {Larson}}, \emph {et~al.},\ }\href@noop {} {\bibfield  {journal} {\bibinfo  {journal} {Physical review letters}\ }\textbf {\bibinfo {volume} {127}},\ \bibinfo {pages} {255101} (\bibinfo {year} {2021})}\BibitemShut {NoStop}%
\bibitem [{\citenamefont {Janssen}\ \emph {et~al.}(2021)\citenamefont {Janssen}, \citenamefont {Falcke}, \citenamefont {Kadler}, \citenamefont {Ros}, \citenamefont {Wielgus}, \citenamefont {Akiyama}, \citenamefont {Balokovi{\'c}}, \citenamefont {Blackburn}, \citenamefont {Bouman}, \citenamefont {Chael} \emph {et~al.}}]{janssen2021event}%
  \BibitemOpen
  \bibfield  {author} {\bibinfo {author} {\bibfnamefont {M.}~\bibnamefont {Janssen}}, \bibinfo {author} {\bibfnamefont {H.}~\bibnamefont {Falcke}}, \bibinfo {author} {\bibfnamefont {M.}~\bibnamefont {Kadler}}, \bibinfo {author} {\bibfnamefont {E.}~\bibnamefont {Ros}}, \bibinfo {author} {\bibfnamefont {M.}~\bibnamefont {Wielgus}}, \bibinfo {author} {\bibfnamefont {K.}~\bibnamefont {Akiyama}}, \bibinfo {author} {\bibfnamefont {M.}~\bibnamefont {Balokovi{\'c}}}, \bibinfo {author} {\bibfnamefont {L.}~\bibnamefont {Blackburn}}, \bibinfo {author} {\bibfnamefont {K.~L.}\ \bibnamefont {Bouman}}, \bibinfo {author} {\bibfnamefont {A.}~\bibnamefont {Chael}}, \emph {et~al.},\ }\href@noop {} {\bibfield  {journal} {\bibinfo  {journal} {Nature Astronomy}\ }\textbf {\bibinfo {volume} {5}},\ \bibinfo {pages} {1017} (\bibinfo {year} {2021})}\BibitemShut {NoStop}%
\bibitem [{\citenamefont {Davelaar}\ \emph {et~al.}(2023)\citenamefont {Davelaar}, \citenamefont {Ripperda}, \citenamefont {Sironi}, \citenamefont {Philippov}, \citenamefont {Olivares}, \citenamefont {Porth}, \citenamefont {Berg}, \citenamefont {Bronzwaer}, \citenamefont {Chatterjee},\ and\ \citenamefont {Liska}}]{davelaar2023synchrotron}%
  \BibitemOpen
  \bibfield  {author} {\bibinfo {author} {\bibfnamefont {J.}~\bibnamefont {Davelaar}}, \bibinfo {author} {\bibfnamefont {B.}~\bibnamefont {Ripperda}}, \bibinfo {author} {\bibfnamefont {L.}~\bibnamefont {Sironi}}, \bibinfo {author} {\bibfnamefont {A.~A.}\ \bibnamefont {Philippov}}, \bibinfo {author} {\bibfnamefont {H.}~\bibnamefont {Olivares}}, \bibinfo {author} {\bibfnamefont {O.}~\bibnamefont {Porth}}, \bibinfo {author} {\bibfnamefont {B.~v.~d.}\ \bibnamefont {Berg}}, \bibinfo {author} {\bibfnamefont {T.}~\bibnamefont {Bronzwaer}}, \bibinfo {author} {\bibfnamefont {K.}~\bibnamefont {Chatterjee}},\ and\ \bibinfo {author} {\bibfnamefont {M.}~\bibnamefont {Liska}},\ }\href@noop {} {\bibfield  {journal} {\bibinfo  {journal} {The Astrophysical Journal Letters}\ }\textbf {\bibinfo {volume} {959}},\ \bibinfo {pages} {L3} (\bibinfo {year} {2023})}\BibitemShut {NoStop}%
\bibitem [{\citenamefont {Guo}\ \emph {et~al.}(2014)\citenamefont {Guo}, \citenamefont {Li}, \citenamefont {Daughton},\ and\ \citenamefont {Liu}}]{guo2014formation}%
  \BibitemOpen
  \bibfield  {author} {\bibinfo {author} {\bibfnamefont {F.}~\bibnamefont {Guo}}, \bibinfo {author} {\bibfnamefont {H.}~\bibnamefont {Li}}, \bibinfo {author} {\bibfnamefont {W.}~\bibnamefont {Daughton}},\ and\ \bibinfo {author} {\bibfnamefont {Y.-H.}\ \bibnamefont {Liu}},\ }\href@noop {} {\bibfield  {journal} {\bibinfo  {journal} {Physical Review Letters}\ }\textbf {\bibinfo {volume} {113}},\ \bibinfo {pages} {155005} (\bibinfo {year} {2014})}\BibitemShut {NoStop}%
\bibitem [{\citenamefont {Sironi}\ and\ \citenamefont {Spitkovsky}(2014)}]{sironi2014relativistic}%
  \BibitemOpen
  \bibfield  {author} {\bibinfo {author} {\bibfnamefont {L.}~\bibnamefont {Sironi}}\ and\ \bibinfo {author} {\bibfnamefont {A.}~\bibnamefont {Spitkovsky}},\ }\href@noop {} {\bibfield  {journal} {\bibinfo  {journal} {The Astrophysical Journal Letters}\ }\textbf {\bibinfo {volume} {783}},\ \bibinfo {pages} {L21} (\bibinfo {year} {2014})}\BibitemShut {NoStop}%
\bibitem [{\citenamefont {Zhdankin}\ \emph {et~al.}(2017)\citenamefont {Zhdankin}, \citenamefont {Werner}, \citenamefont {Uzdensky},\ and\ \citenamefont {Begelman}}]{zhdankin2017kinetic}%
  \BibitemOpen
  \bibfield  {author} {\bibinfo {author} {\bibfnamefont {V.}~\bibnamefont {Zhdankin}}, \bibinfo {author} {\bibfnamefont {G.~R.}\ \bibnamefont {Werner}}, \bibinfo {author} {\bibfnamefont {D.~A.}\ \bibnamefont {Uzdensky}},\ and\ \bibinfo {author} {\bibfnamefont {M.~C.}\ \bibnamefont {Begelman}},\ }\href@noop {} {\bibfield  {journal} {\bibinfo  {journal} {Physical Review Letters}\ }\textbf {\bibinfo {volume} {118}},\ \bibinfo {pages} {055103} (\bibinfo {year} {2017})}\BibitemShut {NoStop}%
\bibitem [{\citenamefont {Comisso}\ and\ \citenamefont {Sironi}(2018)}]{comisso2018particle}%
  \BibitemOpen
  \bibfield  {author} {\bibinfo {author} {\bibfnamefont {L.}~\bibnamefont {Comisso}}\ and\ \bibinfo {author} {\bibfnamefont {L.}~\bibnamefont {Sironi}},\ }\href@noop {} {\bibfield  {journal} {\bibinfo  {journal} {Physical review letters}\ }\textbf {\bibinfo {volume} {121}},\ \bibinfo {pages} {255101} (\bibinfo {year} {2018})}\BibitemShut {NoStop}%
\bibitem [{\citenamefont {Ball}\ \emph {et~al.}(2018)\citenamefont {Ball}, \citenamefont {Sironi},\ and\ \citenamefont {{\"O}zel}}]{ball2018electron}%
  \BibitemOpen
  \bibfield  {author} {\bibinfo {author} {\bibfnamefont {D.}~\bibnamefont {Ball}}, \bibinfo {author} {\bibfnamefont {L.}~\bibnamefont {Sironi}},\ and\ \bibinfo {author} {\bibfnamefont {F.}~\bibnamefont {{\"O}zel}},\ }\href@noop {} {\bibfield  {journal} {\bibinfo  {journal} {The Astrophysical Journal}\ }\textbf {\bibinfo {volume} {862}},\ \bibinfo {pages} {80} (\bibinfo {year} {2018})}\BibitemShut {NoStop}%
\bibitem [{\citenamefont {Comisso}\ and\ \citenamefont {Sironi}(2022)}]{comisso2022ion}%
  \BibitemOpen
  \bibfield  {author} {\bibinfo {author} {\bibfnamefont {L.}~\bibnamefont {Comisso}}\ and\ \bibinfo {author} {\bibfnamefont {L.}~\bibnamefont {Sironi}},\ }\href@noop {} {\bibfield  {journal} {\bibinfo  {journal} {The Astrophysical Journal Letters}\ }\textbf {\bibinfo {volume} {936}},\ \bibinfo {pages} {L27} (\bibinfo {year} {2022})}\BibitemShut {NoStop}%
\bibitem [{\citenamefont {Hellinger}\ and\ \citenamefont {Tr{\'a}vn{\'\i}{\v{c}}ek}(2005)}]{hellinger2005magnetosheath}%
  \BibitemOpen
  \bibfield  {author} {\bibinfo {author} {\bibfnamefont {P.}~\bibnamefont {Hellinger}}\ and\ \bibinfo {author} {\bibfnamefont {P.}~\bibnamefont {Tr{\'a}vn{\'\i}{\v{c}}ek}},\ }\href@noop {} {\bibfield  {journal} {\bibinfo  {journal} {Journal of Geophysical Research: Space Physics}\ }\textbf {\bibinfo {volume} {110}} (\bibinfo {year} {2005})}\BibitemShut {NoStop}%
\bibitem [{\citenamefont {Tr{\'a}vn{\'\i}{\v{c}}ek}\ \emph {et~al.}(2007)\citenamefont {Tr{\'a}vn{\'\i}{\v{c}}ek}, \citenamefont {Hellinger}, \citenamefont {Taylor}, \citenamefont {Escoubet}, \citenamefont {Dandouras},\ and\ \citenamefont {Lucek}}]{travnivcek2007magnetosheath}%
  \BibitemOpen
  \bibfield  {author} {\bibinfo {author} {\bibfnamefont {P.}~\bibnamefont {Tr{\'a}vn{\'\i}{\v{c}}ek}}, \bibinfo {author} {\bibfnamefont {P.}~\bibnamefont {Hellinger}}, \bibinfo {author} {\bibfnamefont {M.~G.}\ \bibnamefont {Taylor}}, \bibinfo {author} {\bibfnamefont {C.~P.}\ \bibnamefont {Escoubet}}, \bibinfo {author} {\bibfnamefont {I.}~\bibnamefont {Dandouras}},\ and\ \bibinfo {author} {\bibfnamefont {E.}~\bibnamefont {Lucek}},\ }\href@noop {} {\bibfield  {journal} {\bibinfo  {journal} {Geophysical research letters}\ }\textbf {\bibinfo {volume} {34}} (\bibinfo {year} {2007})}\BibitemShut {NoStop}%
\bibitem [{\citenamefont {Penna}\ \emph {et~al.}(2010)\citenamefont {Penna}, \citenamefont {McKinney}, \citenamefont {Narayan}, \citenamefont {Tchekhovskoy}, \citenamefont {Shafee},\ and\ \citenamefont {McClintock}}]{penna2010simulations}%
  \BibitemOpen
  \bibfield  {author} {\bibinfo {author} {\bibfnamefont {R.~F.}\ \bibnamefont {Penna}}, \bibinfo {author} {\bibfnamefont {J.~C.}\ \bibnamefont {McKinney}}, \bibinfo {author} {\bibfnamefont {R.}~\bibnamefont {Narayan}}, \bibinfo {author} {\bibfnamefont {A.}~\bibnamefont {Tchekhovskoy}}, \bibinfo {author} {\bibfnamefont {R.}~\bibnamefont {Shafee}},\ and\ \bibinfo {author} {\bibfnamefont {J.~E.}\ \bibnamefont {McClintock}},\ }\href@noop {} {\bibfield  {journal} {\bibinfo  {journal} {Monthly Notices of the Royal Astronomical Society}\ }\textbf {\bibinfo {volume} {408}},\ \bibinfo {pages} {752} (\bibinfo {year} {2010})}\BibitemShut {NoStop}%
\bibitem [{\citenamefont {Ripperda}\ \emph {et~al.}(2020)\citenamefont {Ripperda}, \citenamefont {Bacchini},\ and\ \citenamefont {Philippov}}]{ripperda2020magnetic}%
  \BibitemOpen
  \bibfield  {author} {\bibinfo {author} {\bibfnamefont {B.}~\bibnamefont {Ripperda}}, \bibinfo {author} {\bibfnamefont {F.}~\bibnamefont {Bacchini}},\ and\ \bibinfo {author} {\bibfnamefont {A.~A.}\ \bibnamefont {Philippov}},\ }\href@noop {} {\bibfield  {journal} {\bibinfo  {journal} {The Astrophysical Journal}\ }\textbf {\bibinfo {volume} {900}},\ \bibinfo {pages} {100} (\bibinfo {year} {2020})}\BibitemShut {NoStop}%
\bibitem [{\citenamefont {Kunz}\ \emph {et~al.}(2011)\citenamefont {Kunz}, \citenamefont {Schekochihin}, \citenamefont {Cowley}, \citenamefont {Binney},\ and\ \citenamefont {Sanders}}]{kunz2011thermally}%
  \BibitemOpen
  \bibfield  {author} {\bibinfo {author} {\bibfnamefont {M.}~\bibnamefont {Kunz}}, \bibinfo {author} {\bibfnamefont {A.}~\bibnamefont {Schekochihin}}, \bibinfo {author} {\bibfnamefont {S.}~\bibnamefont {Cowley}}, \bibinfo {author} {\bibfnamefont {J.}~\bibnamefont {Binney}},\ and\ \bibinfo {author} {\bibfnamefont {J.}~\bibnamefont {Sanders}},\ }\href@noop {} {\bibfield  {journal} {\bibinfo  {journal} {Monthly Notices of the Royal Astronomical Society}\ }\textbf {\bibinfo {volume} {410}},\ \bibinfo {pages} {2446} (\bibinfo {year} {2011})}\BibitemShut {NoStop}%
\bibitem [{\citenamefont {Tran}\ \emph {et~al.}(2023)\citenamefont {Tran}, \citenamefont {Sironi}, \citenamefont {Ley}, \citenamefont {Zweibel},\ and\ \citenamefont {Riquelme}}]{tran2023electron}%
  \BibitemOpen
  \bibfield  {author} {\bibinfo {author} {\bibfnamefont {A.}~\bibnamefont {Tran}}, \bibinfo {author} {\bibfnamefont {L.}~\bibnamefont {Sironi}}, \bibinfo {author} {\bibfnamefont {F.}~\bibnamefont {Ley}}, \bibinfo {author} {\bibfnamefont {E.~G.}\ \bibnamefont {Zweibel}},\ and\ \bibinfo {author} {\bibfnamefont {M.~A.}\ \bibnamefont {Riquelme}},\ }\href@noop {} {\bibfield  {journal} {\bibinfo  {journal} {The Astrophysical Journal}\ }\textbf {\bibinfo {volume} {948}},\ \bibinfo {pages} {130} (\bibinfo {year} {2023})}\BibitemShut {NoStop}%
\bibitem [{\citenamefont {Gary}(1991)}]{gary1991electromagnetic}%
  \BibitemOpen
  \bibfield  {author} {\bibinfo {author} {\bibfnamefont {S.~P.}\ \bibnamefont {Gary}},\ }\href@noop {} {\bibfield  {journal} {\bibinfo  {journal} {Space Science Reviews}\ }\textbf {\bibinfo {volume} {56}},\ \bibinfo {pages} {373} (\bibinfo {year} {1991})}\BibitemShut {NoStop}%
\bibitem [{\citenamefont {Gary}(1993)}]{gary1993theory}%
  \BibitemOpen
  \bibfield  {author} {\bibinfo {author} {\bibfnamefont {S.~P.}\ \bibnamefont {Gary}},\ }\href@noop {} {\emph {\bibinfo {title} {Theory of space plasma microinstabilities}}},\ \bibinfo {number} {7}\ (\bibinfo  {publisher} {Cambridge university press},\ \bibinfo {year} {1993})\BibitemShut {NoStop}%
\bibitem [{\citenamefont {Kunz}\ \emph {et~al.}(2014)\citenamefont {Kunz}, \citenamefont {Schekochihin},\ and\ \citenamefont {Stone}}]{kunz2014firehose}%
  \BibitemOpen
  \bibfield  {author} {\bibinfo {author} {\bibfnamefont {M.~W.}\ \bibnamefont {Kunz}}, \bibinfo {author} {\bibfnamefont {A.~A.}\ \bibnamefont {Schekochihin}},\ and\ \bibinfo {author} {\bibfnamefont {J.~M.}\ \bibnamefont {Stone}},\ }\href@noop {} {\bibfield  {journal} {\bibinfo  {journal} {Physical Review Letters}\ }\textbf {\bibinfo {volume} {112}},\ \bibinfo {pages} {205003} (\bibinfo {year} {2014})}\BibitemShut {NoStop}%
\bibitem [{\citenamefont {Berger}\ \emph {et~al.}(1958)\citenamefont {Berger}, \citenamefont {Newcomb}, \citenamefont {Dawson}, \citenamefont {Frieman}, \citenamefont {Kulsrud},\ and\ \citenamefont {Lenard}}]{berger1958heating}%
  \BibitemOpen
  \bibfield  {author} {\bibinfo {author} {\bibfnamefont {J.~M.}\ \bibnamefont {Berger}}, \bibinfo {author} {\bibfnamefont {W.}~\bibnamefont {Newcomb}}, \bibinfo {author} {\bibfnamefont {J.}~\bibnamefont {Dawson}}, \bibinfo {author} {\bibfnamefont {E.}~\bibnamefont {Frieman}}, \bibinfo {author} {\bibfnamefont {R.}~\bibnamefont {Kulsrud}},\ and\ \bibinfo {author} {\bibfnamefont {A.}~\bibnamefont {Lenard}},\ }\href@noop {} {\bibfield  {journal} {\bibinfo  {journal} {The Physics of Fluids}\ }\textbf {\bibinfo {volume} {1}},\ \bibinfo {pages} {301} (\bibinfo {year} {1958})}\BibitemShut {NoStop}%
\bibitem [{\citenamefont {Lichko}\ \emph {et~al.}(2017)\citenamefont {Lichko}, \citenamefont {Egedal}, \citenamefont {Daughton},\ and\ \citenamefont {Kasper}}]{lichko2017magnetic}%
  \BibitemOpen
  \bibfield  {author} {\bibinfo {author} {\bibfnamefont {E.}~\bibnamefont {Lichko}}, \bibinfo {author} {\bibfnamefont {J.}~\bibnamefont {Egedal}}, \bibinfo {author} {\bibfnamefont {W.}~\bibnamefont {Daughton}},\ and\ \bibinfo {author} {\bibfnamefont {J.}~\bibnamefont {Kasper}},\ }\href@noop {} {\bibfield  {journal} {\bibinfo  {journal} {The Astrophysical Journal Letters}\ }\textbf {\bibinfo {volume} {850}},\ \bibinfo {pages} {L28} (\bibinfo {year} {2017})}\BibitemShut {NoStop}%
\bibitem [{\citenamefont {Egedal}\ and\ \citenamefont {Lichko}(2021)}]{egedal2021fast}%
  \BibitemOpen
  \bibfield  {author} {\bibinfo {author} {\bibfnamefont {J.}~\bibnamefont {Egedal}}\ and\ \bibinfo {author} {\bibfnamefont {E.}~\bibnamefont {Lichko}},\ }\href@noop {} {\bibfield  {journal} {\bibinfo  {journal} {Journal of Plasma Physics}\ }\textbf {\bibinfo {volume} {87}},\ \bibinfo {pages} {905870610} (\bibinfo {year} {2021})}\BibitemShut {NoStop}%
\bibitem [{\citenamefont {Montag}\ and\ \citenamefont {Howes}(2022)}]{montag2022field}%
  \BibitemOpen
  \bibfield  {author} {\bibinfo {author} {\bibfnamefont {P.}~\bibnamefont {Montag}}\ and\ \bibinfo {author} {\bibfnamefont {G.~G.}\ \bibnamefont {Howes}},\ }\href@noop {} {\bibfield  {journal} {\bibinfo  {journal} {Physics of plasmas}\ }\textbf {\bibinfo {volume} {29}} (\bibinfo {year} {2022})}\BibitemShut {NoStop}%
\bibitem [{\citenamefont {Malkov}\ and\ \citenamefont {Jebaraj}(2026)}]{malkov2026magnetic}%
  \BibitemOpen
  \bibfield  {author} {\bibinfo {author} {\bibfnamefont {M.}~\bibnamefont {Malkov}}\ and\ \bibinfo {author} {\bibfnamefont {I.}~\bibnamefont {Jebaraj}},\ }\href@noop {} {\bibfield  {journal} {\bibinfo  {journal} {arXiv preprint arXiv:2601.09807}\ } (\bibinfo {year} {2026})}\BibitemShut {NoStop}%
\bibitem [{\citenamefont {Lichko}\ and\ \citenamefont {Egedal}(2020)}]{lichko2020magnetic}%
  \BibitemOpen
  \bibfield  {author} {\bibinfo {author} {\bibfnamefont {E.}~\bibnamefont {Lichko}}\ and\ \bibinfo {author} {\bibfnamefont {J.}~\bibnamefont {Egedal}},\ }\href@noop {} {\bibfield  {journal} {\bibinfo  {journal} {Nature Communications}\ }\textbf {\bibinfo {volume} {11}},\ \bibinfo {pages} {2942} (\bibinfo {year} {2020})}\BibitemShut {NoStop}%
\bibitem [{\citenamefont {Ley}\ \emph {et~al.}(2023)\citenamefont {Ley}, \citenamefont {Zweibel}, \citenamefont {Riquelme}, \citenamefont {Sironi}, \citenamefont {Miller},\ and\ \citenamefont {Tran}}]{ley2023heating}%
  \BibitemOpen
  \bibfield  {author} {\bibinfo {author} {\bibfnamefont {F.}~\bibnamefont {Ley}}, \bibinfo {author} {\bibfnamefont {E.~G.}\ \bibnamefont {Zweibel}}, \bibinfo {author} {\bibfnamefont {M.}~\bibnamefont {Riquelme}}, \bibinfo {author} {\bibfnamefont {L.}~\bibnamefont {Sironi}}, \bibinfo {author} {\bibfnamefont {D.}~\bibnamefont {Miller}},\ and\ \bibinfo {author} {\bibfnamefont {A.}~\bibnamefont {Tran}},\ }\href@noop {} {\bibfield  {journal} {\bibinfo  {journal} {The Astrophysical Journal}\ }\textbf {\bibinfo {volume} {947}},\ \bibinfo {pages} {89} (\bibinfo {year} {2023})}\BibitemShut {NoStop}%
\bibitem [{\citenamefont {Cerutti}\ \emph {et~al.}(2013)\citenamefont {Cerutti}, \citenamefont {Werner}, \citenamefont {Uzdensky},\ and\ \citenamefont {Begelman}}]{cerutti2013simulations}%
  \BibitemOpen
  \bibfield  {author} {\bibinfo {author} {\bibfnamefont {B.}~\bibnamefont {Cerutti}}, \bibinfo {author} {\bibfnamefont {G.~R.}\ \bibnamefont {Werner}}, \bibinfo {author} {\bibfnamefont {D.~A.}\ \bibnamefont {Uzdensky}},\ and\ \bibinfo {author} {\bibfnamefont {M.~C.}\ \bibnamefont {Begelman}},\ }\href@noop {} {\bibfield  {journal} {\bibinfo  {journal} {The Astrophysical Journal}\ }\textbf {\bibinfo {volume} {770}},\ \bibinfo {pages} {147} (\bibinfo {year} {2013})}\BibitemShut {NoStop}%
\bibitem [{\citenamefont {Sironi}\ and\ \citenamefont {Narayan}(2015)}]{sironi2015electron}%
  \BibitemOpen
  \bibfield  {author} {\bibinfo {author} {\bibfnamefont {L.}~\bibnamefont {Sironi}}\ and\ \bibinfo {author} {\bibfnamefont {R.}~\bibnamefont {Narayan}},\ }\href@noop {} {\bibfield  {journal} {\bibinfo  {journal} {The Astrophysical Journal}\ }\textbf {\bibinfo {volume} {800}},\ \bibinfo {pages} {88} (\bibinfo {year} {2015})}\BibitemShut {NoStop}%
\bibitem [{\citenamefont {Bacchini}\ \emph {et~al.}(2026)\citenamefont {Bacchini}, \citenamefont {Gorbunov}, \citenamefont {de~Bonhome}, \citenamefont {Els}, \citenamefont {Argyropoulos}, \citenamefont {Ly},\ and\ \citenamefont {Gro{\v{s}}elj}}]{bacchini2026particle}%
  \BibitemOpen
  \bibfield  {author} {\bibinfo {author} {\bibfnamefont {F.}~\bibnamefont {Bacchini}}, \bibinfo {author} {\bibfnamefont {E.~A.}\ \bibnamefont {Gorbunov}}, \bibinfo {author} {\bibfnamefont {M.~P.}\ \bibnamefont {de~Bonhome}}, \bibinfo {author} {\bibfnamefont {P.}~\bibnamefont {Els}}, \bibinfo {author} {\bibfnamefont {K.-X.}\ \bibnamefont {Argyropoulos}}, \bibinfo {author} {\bibfnamefont {M.~N.}\ \bibnamefont {Ly}},\ and\ \bibinfo {author} {\bibfnamefont {D.}~\bibnamefont {Gro{\v{s}}elj}},\ }\href@noop {} {\bibfield  {journal} {\bibinfo  {journal} {Plasma Physics and Controlled Fusion}\ }\textbf {\bibinfo {volume} {68}},\ \bibinfo {pages} {055031} (\bibinfo {year} {2026})}\BibitemShut {NoStop}%
\bibitem [{\citenamefont {Péters~de Bonhome}\ \emph {et~al.}(2026)\citenamefont {Péters~de Bonhome}, \citenamefont {Bacchini}, \citenamefont {Pezzini},\ and\ \citenamefont {Pierrard}}]{petersdebonhome2026}%
  \BibitemOpen
  \bibfield  {author} {\bibinfo {author} {\bibfnamefont {M.}~\bibnamefont {Péters~de Bonhome}}, \bibinfo {author} {\bibfnamefont {F.}~\bibnamefont {Bacchini}}, \bibinfo {author} {\bibfnamefont {L.}~\bibnamefont {Pezzini}},\ and\ \bibinfo {author} {\bibfnamefont {V.}~\bibnamefont {Pierrard}},\ }\href@noop {} {\bibfield  {journal} {\bibinfo  {journal} {arXiv:2605.01895}\ } (\bibinfo {year} {2026})}\BibitemShut {NoStop}%
\bibitem [{Note1()}]{Note1}%
  \BibitemOpen
  \bibinfo {note} {The Maxwell--J{\"u}ttner distribution is the relativistic generalization of the Maxwell distribution, typically used to initialize particle-in-cell simulations.}\BibitemShut {Stop}%
\bibitem [{\citenamefont {J{\"u}ttner}(1911)}]{juttner1911maxwellsche}%
  \BibitemOpen
  \bibfield  {author} {\bibinfo {author} {\bibfnamefont {F.}~\bibnamefont {J{\"u}ttner}},\ }\href@noop {} {\bibfield  {journal} {\bibinfo  {journal} {Annalen der Physik}\ }\textbf {\bibinfo {volume} {339}},\ \bibinfo {pages} {856} (\bibinfo {year} {1911})}\BibitemShut {NoStop}%
\bibitem [{Note2()}]{Note2}%
  \BibitemOpen
  \bibinfo {note} {We require $\tau _0$ to be larger than time-scales associated with instabilities, comparable to particle gyroperiods.}\BibitemShut {Stop}%
\bibitem [{\citenamefont {Gary}(1992)}]{gary1992mirror}%
  \BibitemOpen
  \bibfield  {author} {\bibinfo {author} {\bibfnamefont {S.~P.}\ \bibnamefont {Gary}},\ }\href@noop {} {\bibfield  {journal} {\bibinfo  {journal} {Journal of Geophysical Research: Space Physics}\ }\textbf {\bibinfo {volume} {97}},\ \bibinfo {pages} {8519} (\bibinfo {year} {1992})}\BibitemShut {NoStop}%
\bibitem [{\citenamefont {Southwood}\ and\ \citenamefont {Kivelson}(1993)}]{southwood1993mirror}%
  \BibitemOpen
  \bibfield  {author} {\bibinfo {author} {\bibfnamefont {D.~J.}\ \bibnamefont {Southwood}}\ and\ \bibinfo {author} {\bibfnamefont {M.~G.}\ \bibnamefont {Kivelson}},\ }\href@noop {} {\bibfield  {journal} {\bibinfo  {journal} {Journal of Geophysical Research: Space Physics}\ }\textbf {\bibinfo {volume} {98}},\ \bibinfo {pages} {9181} (\bibinfo {year} {1993})}\BibitemShut {NoStop}%
\bibitem [{\citenamefont {Gary}\ \emph {et~al.}(1998)\citenamefont {Gary}, \citenamefont {Li}, \citenamefont {O'Rourke},\ and\ \citenamefont {Winske}}]{gary1998proton}%
  \BibitemOpen
  \bibfield  {author} {\bibinfo {author} {\bibfnamefont {S.~P.}\ \bibnamefont {Gary}}, \bibinfo {author} {\bibfnamefont {H.}~\bibnamefont {Li}}, \bibinfo {author} {\bibfnamefont {S.}~\bibnamefont {O'Rourke}},\ and\ \bibinfo {author} {\bibfnamefont {D.}~\bibnamefont {Winske}},\ }\href@noop {} {\bibfield  {journal} {\bibinfo  {journal} {Journal of Geophysical Research: Space Physics}\ }\textbf {\bibinfo {volume} {103}},\ \bibinfo {pages} {14567} (\bibinfo {year} {1998})}\BibitemShut {NoStop}%
\bibitem [{\citenamefont {Zhdankin}\ \emph {et~al.}(2023)\citenamefont {Zhdankin}, \citenamefont {Kunz},\ and\ \citenamefont {Uzdensky}}]{zhdankin2023synchrotron}%
  \BibitemOpen
  \bibfield  {author} {\bibinfo {author} {\bibfnamefont {V.}~\bibnamefont {Zhdankin}}, \bibinfo {author} {\bibfnamefont {M.~W.}\ \bibnamefont {Kunz}},\ and\ \bibinfo {author} {\bibfnamefont {D.~A.}\ \bibnamefont {Uzdensky}},\ }\href@noop {} {\bibfield  {journal} {\bibinfo  {journal} {The Astrophysical Journal}\ }\textbf {\bibinfo {volume} {944}},\ \bibinfo {pages} {24} (\bibinfo {year} {2023})}\BibitemShut {NoStop}%
\bibitem [{\citenamefont {L{\'o}pez}\ \emph {et~al.}(2016)\citenamefont {L{\'o}pez}, \citenamefont {Moya}, \citenamefont {Navarro}, \citenamefont {Araneda}, \citenamefont {Mu{\~n}oz}, \citenamefont {Vi{\~n}as},\ and\ \citenamefont {Valdivia}}]{lopez2016relativistic}%
  \BibitemOpen
  \bibfield  {author} {\bibinfo {author} {\bibfnamefont {R.~A.}\ \bibnamefont {L{\'o}pez}}, \bibinfo {author} {\bibfnamefont {P.~S.}\ \bibnamefont {Moya}}, \bibinfo {author} {\bibfnamefont {R.~E.}\ \bibnamefont {Navarro}}, \bibinfo {author} {\bibfnamefont {J.~A.}\ \bibnamefont {Araneda}}, \bibinfo {author} {\bibfnamefont {V.}~\bibnamefont {Mu{\~n}oz}}, \bibinfo {author} {\bibfnamefont {A.~F.}\ \bibnamefont {Vi{\~n}as}},\ and\ \bibinfo {author} {\bibfnamefont {J.~A.}\ \bibnamefont {Valdivia}},\ }\href@noop {} {\bibfield  {journal} {\bibinfo  {journal} {The Astrophysical Journal}\ }\textbf {\bibinfo {volume} {832}},\ \bibinfo {pages} {36} (\bibinfo {year} {2016})}\BibitemShut {NoStop}%
\bibitem [{SMP()}]{SMPB}%
  \BibitemOpen
  \href@noop {} {}\bibinfo {note} {Supplemental material with additional numerical analysis and simulation animation.}\BibitemShut {Stop}%
\bibitem [{Note3()}]{Note3}%
  \BibitemOpen
  \bibinfo {note} {The corresponding energy distribution is ${\protect \rm d}N/{\protect \rm d}\epsilon \protect \!=\protect \!({\protect \rm d}N/{\protect \rm d}p)({\protect \rm d}p/{\protect \rm d}\epsilon )$, where $\epsilon $ is the kinetic energy.}\BibitemShut {Stop}%
\bibitem [{\citenamefont {Zhdankin}(2022{\natexlab{a}})}]{zhdankin2022non}%
  \BibitemOpen
  \bibfield  {author} {\bibinfo {author} {\bibfnamefont {V.}~\bibnamefont {Zhdankin}},\ }\href@noop {} {\bibfield  {journal} {\bibinfo  {journal} {Journal of Plasma Physics}\ }\textbf {\bibinfo {volume} {88}},\ \bibinfo {pages} {175880303} (\bibinfo {year} {2022}{\natexlab{a}})}\BibitemShut {NoStop}%
\bibitem [{\citenamefont {Comisso}\ \emph {et~al.}(2024)\citenamefont {Comisso}, \citenamefont {Farrar},\ and\ \citenamefont {Muzio}}]{comisso2024ultra}%
  \BibitemOpen
  \bibfield  {author} {\bibinfo {author} {\bibfnamefont {L.}~\bibnamefont {Comisso}}, \bibinfo {author} {\bibfnamefont {G.~R.}\ \bibnamefont {Farrar}},\ and\ \bibinfo {author} {\bibfnamefont {M.~S.}\ \bibnamefont {Muzio}},\ }\href@noop {} {\bibfield  {journal} {\bibinfo  {journal} {The Astrophysical Journal Letters}\ }\textbf {\bibinfo {volume} {977}},\ \bibinfo {pages} {L18} (\bibinfo {year} {2024})}\BibitemShut {NoStop}%
\bibitem [{\citenamefont {Hillas}(1984)}]{hillas1984origin}%
  \BibitemOpen
  \bibfield  {author} {\bibinfo {author} {\bibfnamefont {A.~M.}\ \bibnamefont {Hillas}},\ }\href@noop {} {\bibfield  {journal} {\bibinfo  {journal} {IN: Annual review of astronomy and astrophysics. Volume 22. Palo Alto, CA, Annual Reviews, Inc., 1984, p. 425-444.}\ }\textbf {\bibinfo {volume} {22}},\ \bibinfo {pages} {425} (\bibinfo {year} {1984})}\BibitemShut {NoStop}%
\bibitem [{\citenamefont {Zhdankin}(2022{\natexlab{b}})}]{zhdankin2022generalized}%
  \BibitemOpen
  \bibfield  {author} {\bibinfo {author} {\bibfnamefont {V.}~\bibnamefont {Zhdankin}},\ }\href@noop {} {\bibfield  {journal} {\bibinfo  {journal} {Physical Review X}\ }\textbf {\bibinfo {volume} {12}},\ \bibinfo {pages} {031011} (\bibinfo {year} {2022}{\natexlab{b}})}\BibitemShut {NoStop}%
\bibitem [{\citenamefont {Zhdankin}(2023)}]{zhdankin2023dimensional}%
  \BibitemOpen
  \bibfield  {author} {\bibinfo {author} {\bibfnamefont {V.}~\bibnamefont {Zhdankin}},\ }\href@noop {} {\bibfield  {journal} {\bibinfo  {journal} {Journal of Physics A: Mathematical and Theoretical}\ }\textbf {\bibinfo {volume} {56}},\ \bibinfo {pages} {385002} (\bibinfo {year} {2023})}\BibitemShut {NoStop}%
\bibitem [{Note4()}]{Note4}%
  \BibitemOpen
  \bibinfo {note} {We choose an ultrarelativistic distribution because it is appropriate to describe the tail of electron distributions from simulations, extending into the $p\protect \!\gg \protect \!m_e c$ range, and because it allows for the model to be derived analytically in closed form.}\BibitemShut {Stop}%
\bibitem [{Note5()}]{Note5}%
  \BibitemOpen
  \bibinfo {note} {$\alpha \protect \!\leqslant \protect \!1$ would cause $N$ to diverge in large-scale systems, where $p_c\protect \!\rightarrow \protect \!\infty $.}\BibitemShut {Stop}%
\bibitem [{\citenamefont {Tsallis}(1988)}]{tsallis1988possible}%
  \BibitemOpen
  \bibfield  {author} {\bibinfo {author} {\bibfnamefont {C.}~\bibnamefont {Tsallis}},\ }\href@noop {} {\bibfield  {journal} {\bibinfo  {journal} {Journal of statistical physics}\ }\textbf {\bibinfo {volume} {52}},\ \bibinfo {pages} {479} (\bibinfo {year} {1988})}\BibitemShut {NoStop}%
\bibitem [{\citenamefont {Livadiotis}\ and\ \citenamefont {McComas}(2013)}]{livadiotis2013understanding}%
  \BibitemOpen
  \bibfield  {author} {\bibinfo {author} {\bibfnamefont {G.}~\bibnamefont {Livadiotis}}\ and\ \bibinfo {author} {\bibfnamefont {D.~J.}\ \bibnamefont {McComas}},\ }\href@noop {} {\bibfield  {journal} {\bibinfo  {journal} {Space Science Reviews}\ }\textbf {\bibinfo {volume} {175}},\ \bibinfo {pages} {183} (\bibinfo {year} {2013})}\BibitemShut {NoStop}%
\bibitem [{\citenamefont {Livadiotis}(2018)}]{livadiotis2018thermodynamic}%
  \BibitemOpen
  \bibfield  {author} {\bibinfo {author} {\bibfnamefont {G.}~\bibnamefont {Livadiotis}},\ }\href@noop {} {\bibfield  {journal} {\bibinfo  {journal} {Europhysics Letters}\ }\textbf {\bibinfo {volume} {122}},\ \bibinfo {pages} {50001} (\bibinfo {year} {2018})}\BibitemShut {NoStop}%
\bibitem [{\citenamefont {Livadiotis}\ and\ \citenamefont {McComas}(2021)}]{livadiotis2021thermodynamic}%
  \BibitemOpen
  \bibfield  {author} {\bibinfo {author} {\bibfnamefont {G.}~\bibnamefont {Livadiotis}}\ and\ \bibinfo {author} {\bibfnamefont {D.~J.}\ \bibnamefont {McComas}},\ }\href@noop {} {\bibfield  {journal} {\bibinfo  {journal} {Entropy}\ }\textbf {\bibinfo {volume} {23}},\ \bibinfo {pages} {1683} (\bibinfo {year} {2021})}\BibitemShut {NoStop}%
\bibitem [{\citenamefont {Zhdankin}\ \emph {et~al.}(2019)\citenamefont {Zhdankin}, \citenamefont {Uzdensky}, \citenamefont {Werner},\ and\ \citenamefont {Begelman}}]{zhdankin2019electron}%
  \BibitemOpen
  \bibfield  {author} {\bibinfo {author} {\bibfnamefont {V.}~\bibnamefont {Zhdankin}}, \bibinfo {author} {\bibfnamefont {D.~A.}\ \bibnamefont {Uzdensky}}, \bibinfo {author} {\bibfnamefont {G.~R.}\ \bibnamefont {Werner}},\ and\ \bibinfo {author} {\bibfnamefont {M.~C.}\ \bibnamefont {Begelman}},\ }\href@noop {} {\bibfield  {journal} {\bibinfo  {journal} {Physical review letters}\ }\textbf {\bibinfo {volume} {122}},\ \bibinfo {pages} {055101} (\bibinfo {year} {2019})}\BibitemShut {NoStop}%
\bibitem [{\citenamefont {Arr{\`o}}\ \emph {et~al.}(2022)\citenamefont {Arr{\`o}}, \citenamefont {Califano},\ and\ \citenamefont {Lapenta}}]{arro2022spectral}%
  \BibitemOpen
  \bibfield  {author} {\bibinfo {author} {\bibfnamefont {G.}~\bibnamefont {Arr{\`o}}}, \bibinfo {author} {\bibfnamefont {F.}~\bibnamefont {Califano}},\ and\ \bibinfo {author} {\bibfnamefont {G.}~\bibnamefont {Lapenta}},\ }\href@noop {} {\bibfield  {journal} {\bibinfo  {journal} {Astronomy \& Astrophysics}\ }\textbf {\bibinfo {volume} {668}},\ \bibinfo {pages} {A33} (\bibinfo {year} {2022})}\BibitemShut {NoStop}%
\end{thebibliography}
%

\end{document}